\documentclass[prb,aps,twocolumn,superscriptaddress,floatfix,showpacs,longbibliography]{revtex4-2}
\usepackage{graphicx}
\usepackage{dcolumn}
\usepackage{bm}
\usepackage{upgreek}
\usepackage{natbib}
\usepackage[normalem]{ulem}
\usepackage{amsmath,amssymb}
\usepackage{dsfont}
\usepackage[english]{babel}
\usepackage{color}
\usepackage{lipsum}
\usepackage{autobreak}
\usepackage[colorlinks=true, linkcolor=blue, citecolor=blue, urlcolor=blue]{hyperref}
\usepackage{float}

\begin{document}

\title{$\mathbb{Z}_2$ topological trion insulator}

\author{Yichen Chu}
\affiliation{Guangdong Basic Research Center of Excellence for Structure and Fundamental Interactions of Matter, Guangdong Provincial Key Laboratory of Quantum Engineering and Quantum Materials, School of Physics, South China Normal University, Guangzhou 510006, China}
\affiliation{Guangdong-Hong Kong Joint Laboratory of Quantum Matter, Frontier
Research Institute for Physics, South China Normal University, Guangzhou
510006, China}

\author{Qizhong Zhu}
\email{qzzhu@m.scnu.edu.cn}
\affiliation{Guangdong Basic Research Center of Excellence for Structure and Fundamental Interactions of Matter, Guangdong Provincial Key Laboratory of Quantum Engineering and Quantum Materials, School of Physics, South China Normal University, Guangzhou 510006, China}
\affiliation{Guangdong-Hong Kong Joint Laboratory of Quantum Matter, Frontier
Research Institute for Physics, South China Normal University, Guangzhou
510006, China}

\date{\today}

\begin{abstract}

Trions, charged quasiparticles formed by binding an exciton to an excess charge carrier, dominate the optical response of doped transition metal dichalcogenides (TMDs), and the study of the transport properties of trions in TMDs may have application in developing high-speed excitonic and optoelectronic devices. However, an important building block for low-dissipation optoelectronic devices that provides dissipationless transport channels for trions has remained elusive. Here, we propose the concept of a $\mathbb{Z}_2$ topological trion insulator that features helical dissipationless edge states for trions. This is realized for intralayer trions, which inherit the valley-orbit coupling of intralayer excitons in TMDs subject to a moir\'e periodic potential. We find that under certain circumstances, the moir\'e trion band becomes topological, characterized by the $\mathbb{Z}_2$ topological number. We further provide two specific material realizations of this $\mathbb{Z}_2$ topological insulator: a doped monolayer TMD placed on top of a twisted hBN substrate, and a generic twisted TMD heterobilayer. We also examine the effect of charge screening and find that the $\mathbb{Z}_2$ topological trion insulator remains robust. Our work paves the way toward realizing dissipationless excitonic and trionic devices.

\end{abstract}

{\maketitle}

\section{Introduction}
In monolayer transition metal dichalcogenides (TMDs), strong Coulomb interactions bind electrons and holes together to form excitons \cite{heTightlyBoundExcitons2014,sallenRobustOpticalEmission2012,kornLowtemperaturePhotocarrierDynamics2011,ugedaGiantBandgapRenormalization2014,wangGiantEnhancementOptical2015,chernikovExcitonBindingEnergy2014,qiuOpticalSpectrumOfMoS22013,komsaEffectsConfinementEnvironment2012}. The large binding energy and unique optical selection rules of valley excitons attracted significant attention \cite{hanbickiMeasurementHighExciton2015,chernikovExcitonBindingEnergy2014,hsuDielectricImpactExciton2019,ugedaGiantBandgapRenormalization2014,parkDirectDeterminationMonolayer2018,kylanpaaBindingEnergiesExciton2015,zhuExcitonBindingEnergy2015,yuAnomalousLightCones2015,liuOpticalPropertiesMonolayer2014,heTightlyBoundExcitons2014,yeProbingExcitonicDark2014,xiaoNonlinearOpticalSelection2015}. When the system is doped with extra charge carriers, a bound electron-hole pair can capture an excess electron (hole), forming a negatively (positively) charged quasiparticle known as a trion \cite{courtadeChargedExcitonsMonolayer2017,vaclavkovaSingletTripletTrions2018,jadczakProbingFreeLocalized2017,jakubczykImpactEnvironmentDynamics2018,szyniszewskiBindingEnergiesTrions2017,vanderdonckExcitonsTrionsMonolayer2017}, which also inherits the valley degrees of freedom \cite{berkelbachTheoryNeutralCharged2013,makTightlyBoundTrions2013,luiTrionInducedNegativePhotoconductivity2014,filinovInfluenceWellwidthFluctuations2004}. 
Compared with charge neutral exciton, trions can be more conveniently manipulated with external electric field. The study of trions in TMDs is essential for unlocking the full potential of these atomically thin semiconductors in next-generation optoelectronic and quantum devices. 
In recent years, numerous theoretical and experimental studies have investigated the transport properties of trions in TMDs, including the trion Hall effect \cite{yuDiracConesDirac2014,hichriTrionFineStructure2020}, trion-phonon interactions \cite{perea-causinTrionphononInteractionAtomically2022,meierEmergentTrionphononCoupling2023}, trion formation dynamics \cite{singhTrionFormationDynamics2016}, valley depolarization \cite{singhLongLivedValleyPolarization2016,anStrainControlExciton2023,plechingerTrionFineStructure2016} and trion diffusion \cite{beretNonlinearDiffusionNegatively2023,chengObservationDiffusionDrift2021,kimFreeTrionsUnity2022}. 
Despite these efforts, whether one can realize the dissipationless transport of trions remains an open question.

Notably, intralayer excitons in monolayer TMDs have a key feature known as valley-orbit coupling, which describes the coupling between valley degrees of freedom with center-of-mass momentum of excitons, due to the electron-hole exchange interaction \cite{yuDiracConesDirac2014,qiuNonanalyticityValleyQuantum2015,wuExcitonBandStructure2015}. This valley-orbit coupling leads to a variety of interesting phenomena, including exciton linear dispersion \cite{yuDiracConesDirac2014,qiuNonanalyticityValleyQuantum2015,wuExcitonBandStructure2015,shimazakiStronglyCorrelatedElectrons2020,simbulanSelectivePhotoexcitationFiniteMomentum2021}, valley depolarization \cite{yuValleyDepolarizationDue2014,yangExcitonValleyDepolarization2020,yuValleyDepolarizationDynamics2016}, topological exciton band \cite{wuTopologicalExcitonBands2017}, chiral excitonics \cite{yangChiralExcitonicsMonolayer2022}, single-photon emitters with polarization and angular momentum locking \cite{zhangSinglePhotonEmitters2023} and possible unconventional superfluid \cite{chenSearchingUnconventionalSuperfluid2023}.

In this work, we propose a novel state of matter,
i.e., the $\mathbb{Z}_2$ topological trion insulator, which is a topological insulator for trions with time-reversal symmetry, as a mechanism of realizing dissipationless trion transport. We first present the theoretical framework for realizing the $\mathbb{Z}_2$ topological trion insulator, which is made possible by the valley-orbit coupling of intralayer excitons in TMD, and then describe two specific material systems where this state can be realized. One realization involves placing a monolayer TMD on a twisted hBN substrate, where electric polarization is generated at the interface between the twisted hBN layers, creating an electrostatic moir\'e potential that acts on the TMD monolayer. The other realization utilizes the twisted TMD heterobilayers, where the twisting naturally create a moir\'e potential for the intralayer trions. With electric doping of excess electrons preserving valley symmetry, we find that the lowest moir\'e trion bands become topological within a certain range of twist angles, with opposite Chern numbers in different spin sectors. This topological trion insulator features helical edge states, similar to $\mathbb{Z}_2$ topological insulator for electrons \cite{kaneQuantumSpinHall2005,kaneTopologicalOrderQuantum2005}. We also examine the effect of Coulomb screening and verifies the stability of the topological trion insulator.

The rest of the paper is organized as follows. In Sec. \ref{sec2}, we present the general framework for realizing a topological trion insulator, focusing on the roles of electron-hole exchange interaction and moiré potential. In Sec. \ref{sec3}, we examine two material systems that can host a topological trion insulator and provide the corresponding topological phase diagram. We also discuss the impact of electrostatic screening. Finally, in Sec. \ref{sec4}, we address the experimental observation of our predictions and conclude the paper.

\section{Model Hamiltonian of the $\mathbb{Z}_2$  topological trion insulator }
\label{sec2}

Intralayer excitons in the two valleys of TMDs are not independent; instead, they are coupled through the electron-hole exchange interaction between the valleys. This exchange interaction leads to interband transitions of electrons between the valleys, resulting in a flip of the exciton valley pseudospin. Because of the electron-hole exchange interaction, the center-of-mass momentum of intralayer excitons in monolayer TMDs is coupled with the valley pseudospin, which results in valley-orbit coupling. The effective Hamiltonian of the valley excitons with valley-orbit coupling reads follows \cite{wuTopologicalExcitonBands2017,yuDiracConesDirac2014,qiuNonanalyticityValleyQuantum2015,wuExcitonBandStructure2015,sauerOpticalEmissionLightlike2021,salvadorOpticalSignaturesPeriodic2022}
\begin{equation}
	\begin{split}
		\hat{H}_{\text{ex }}&=\frac{\hbar^{2}\boldsymbol{Q}^{2}}{2m_{\text{ex}}}+\beta \boldsymbol{Q}+\beta \boldsymbol{Q}\cos(2\boldsymbol{\phi}_{\boldsymbol{\boldsymbol{Q}}})\sigma_{x}\\
		&+\beta \boldsymbol{Q}\sin(2\boldsymbol{\phi}_{\boldsymbol{\boldsymbol{Q}}})\sigma_{y}. 
	\end{split}
\end{equation}
Here $\boldsymbol{Q}$ represents the center-of-mass momentum of the exciton, $m_{\text{ex}}$ is its effective mass, $\beta\approx0.9$ eV$\cdot${\AA} stands for the valley-orbit coupling strength \cite{qiuNonanalyticityValleyQuantum2015} and $\sigma_i$ is the Pauli matrix defined in the valley pseudospin space. The Hamiltonian describes an exciton dispersion with two branches, i.e., a parabolic lower branch and a ``V''-shaped upper branch, which meet at the $\mathbf{\gamma}$ point. When time-reversal symmetry is broken, for example by an external magnetic field, an additional $\delta\sigma_z$ term appears in the Hamiltonian, leading to a gap opening at the $\mathbf{\gamma}$ point.

Valley-orbit coupling, similar to conventional spin-orbit coupling of electrons, creates a spin texture in momentum space and can lead to possible topological phases of excitons. For instance, in twisted heterobilayers, intralayer excitons experience an additional moir\'e potential. It has been found that valley-orbit coupled excitons in a moir\'e potential can give rise to topological exciton bands when time-reversal symmetry is broken by an external magnetic field or valley-selective optical Stark effect \cite{wuTopologicalExcitonBands2017}.

\begin{figure}[h]
 	\centering 
 	\includegraphics[width=0.95\linewidth]{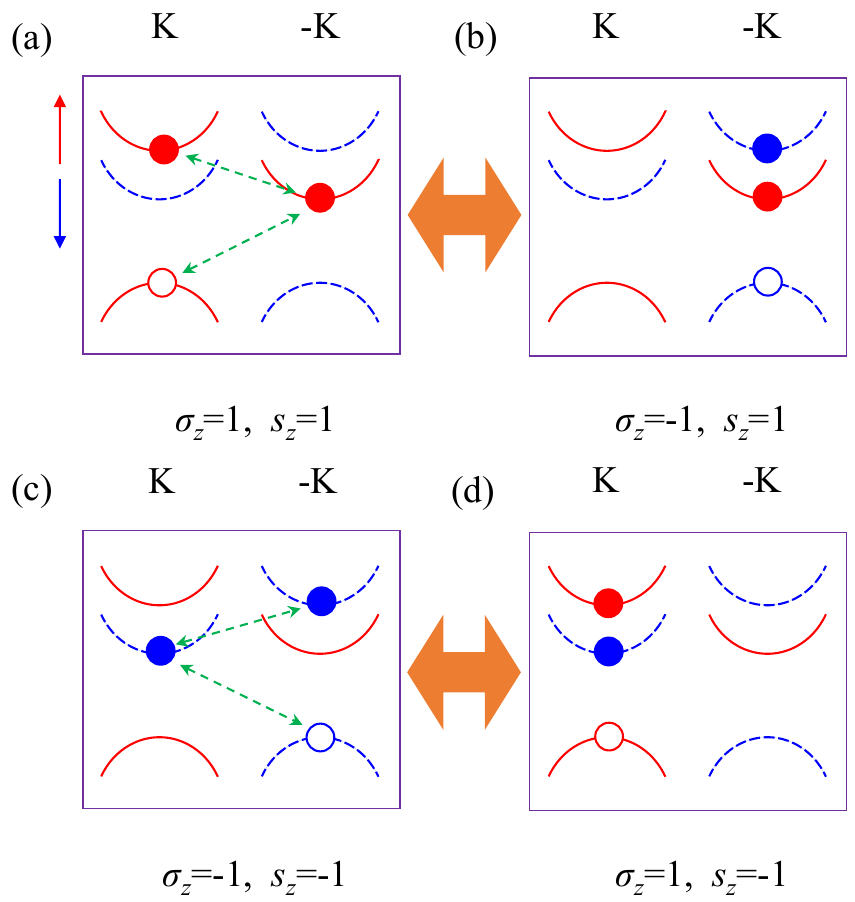}	
 	\caption{(a)-(d) Schematic representations of the four configurations of negatively charged trions in a TMD monolayer. Red solid curves denote spin-up conduction and valence bands, while blue dashed curves indicate spin-down bands. Solid dots represent electrons; open circles represent holes. Dashed double arrows illustrate the exchange interaction between the excess electron and the electron-hole pair, present only in configurations (a) and (c). Solid double arrows depict the electron–hole exchange interaction that couples excitons in different valleys.} 
 	\label{<fig_four_configurations>}%
 \end{figure}

When the system is doped with extra electrons or holes, e.g., by tuning the gate voltage, charged excitons, known as trions, are formed. Trions carry the same charge as the excess electron or hole bound to the exciton, allowing for convenient electrical manipulation using an external electric field. In the following, we focus on the negatively charged trion. Due to spin-orbit coupling, the excess electron in the lower conduction band exhibits spin-valley locking.
In this system, excitons have two valley pseudospins, and the excess electrons can also occupy two valleys or different spin states, resulting in four possible trion configurations, as shown in Fig. \ref{<fig_four_configurations>}. The exchange interaction between the excess electrons and the electron-hole pairs acts like a Zeeman energy, splitting the four trion configurations into two  separate energy pairs. We use $\sigma_{z} = \pm 1$ to label the valley pseudospin of excitons and $s_{z} = \pm 1$ to label the spin of the excess electron. The effect of the electron-hole exchange interaction between the excess electron and the electron-hole pair can be described by the term $\frac{\delta}{2}(\sigma_{z}s_{z} + 1)$, where $\delta$ is the strength of the energy splitting caused by the exchange interaction \cite{yuDiracConesDirac2014,yuValleyExcitonsTwodimensional2015}. For a given spin state of the excess electron, the Zeeman energy induced by the electron-hole exchange interaction naturally breaks time-reversal symmetry for valley excitons. This opens a gap at the $\mathbf{\gamma}$ point and leads to the formation of nonzero Chern numbers in the trion bands.
 The sign of the valley Zeeman energy, or the direction of the effective magnetic field, depends on the spin of the excess electron. When electrons are doped through an electrical gate, time-reversal symmetry in the system is restored by ensuring equal electron densities in both spin states. 

Taking into account both the electron-hole exchange interaction and the exchange interaction between the excess electron and exciton, the trion Hamiltonian is given by \cite{yuDiracConesDirac2014}
\begin{equation}
	\begin{split}
\hat{H}_{\text{trion}}&=\frac{\hbar^{2}\boldsymbol{Q}^{2}}{2m_{\text{trion}}}+\beta \boldsymbol{Q}+\beta \boldsymbol{Q}\cos(2\boldsymbol{\phi}_{\boldsymbol{\boldsymbol{Q}}})\sigma_{x}\\
		&+\beta \boldsymbol{Q}\sin(2\boldsymbol{\phi}_{\boldsymbol{\boldsymbol{Q}}})\sigma_{y}+\frac{\delta}{2}(\sigma_{z}s_{z}+1), 
	\end{split}
\end{equation}
where $m_{\text{trion}}$ denotes the trion effective mass. To realize a topological trion band, we consider the situation where trions are placed in an external periodic potential. This can be easily achieved using an electrostatic potential, since trions are charged quasiparticles.

When subject to an external periodic potential, the trion dispersion in each spin state forms bands. For excess electron with spin $s_z=\pm1$, the Chern number of $m$-th trion band is calculated by integrating the Berry curvature over the first Brillouin zone, i.e.,
 \begin{equation}
\mathcal{C}_{m}^{\pm}=\frac{1}{2\pi}\int_{\mathrm{BZ}} d^{2}\mathbf{k}\mathcal{F}^{\pm}_{m}({\mathbf{k}}),
 	\label{<eq_berry>}%
 \end{equation}
 where $\mathbf{k}$ is the quasi-momentum, $u^{\pm}_m(\mathbf{k})$ is the cell-periodic part of Bloch wave function in the $m$-th band for spin $s_z=\pm1$, and $\mathcal{F}^{\pm}_{m}({\mathbf{k}})=i\left\langle \nabla_{\mathbf{k}}u^{\pm}_m(\mathbf{k})\right|\times\left|\nabla_{\mathbf{k}}u^{\pm}_m(\mathbf{k})\right\rangle$ is the corresponding Berry curvature. It is found that $\mathcal{F}^{+}_{m}({\mathbf{k}})=-\mathcal{F}^{-}_{m}({\mathbf{k}})$, similar to gapped Dirac model, so the Chern numbers of the trion band for different spin states are opposite. Although the overall Chern number is zero, we can still use the $\mathbb{Z}_2$ topological number to characterize the topological properties of the system. In particular, with the spin of excess electron being a good quantum number, the topological properties of the system can be described by the spin Chern number, i.e., $\mathcal{C}_{\mathrm{spin}}=(\mathcal{C}^+-\mathcal{C}^-)/2$. In this case,
the $\mathbb{Z}_2$ number $\nu$ is simply given by, $\nu=\mathcal{C}_{\mathrm{spin}}~\mathrm{mod}~2$ \cite{hasanColloquiumTopologicalInsulators2010}.  Below we give two exemplary realizations of the $\mathbb{Z}_2$ topological trion insulator in moir\'e superlattices.
 
\section{Possible material realizations}
\label{sec3}
\subsection{Monolayer TMD on twisted hBN substrate}
\label{model1}

\begin{figure}
	\centering 
	\includegraphics[width=0.95\linewidth]{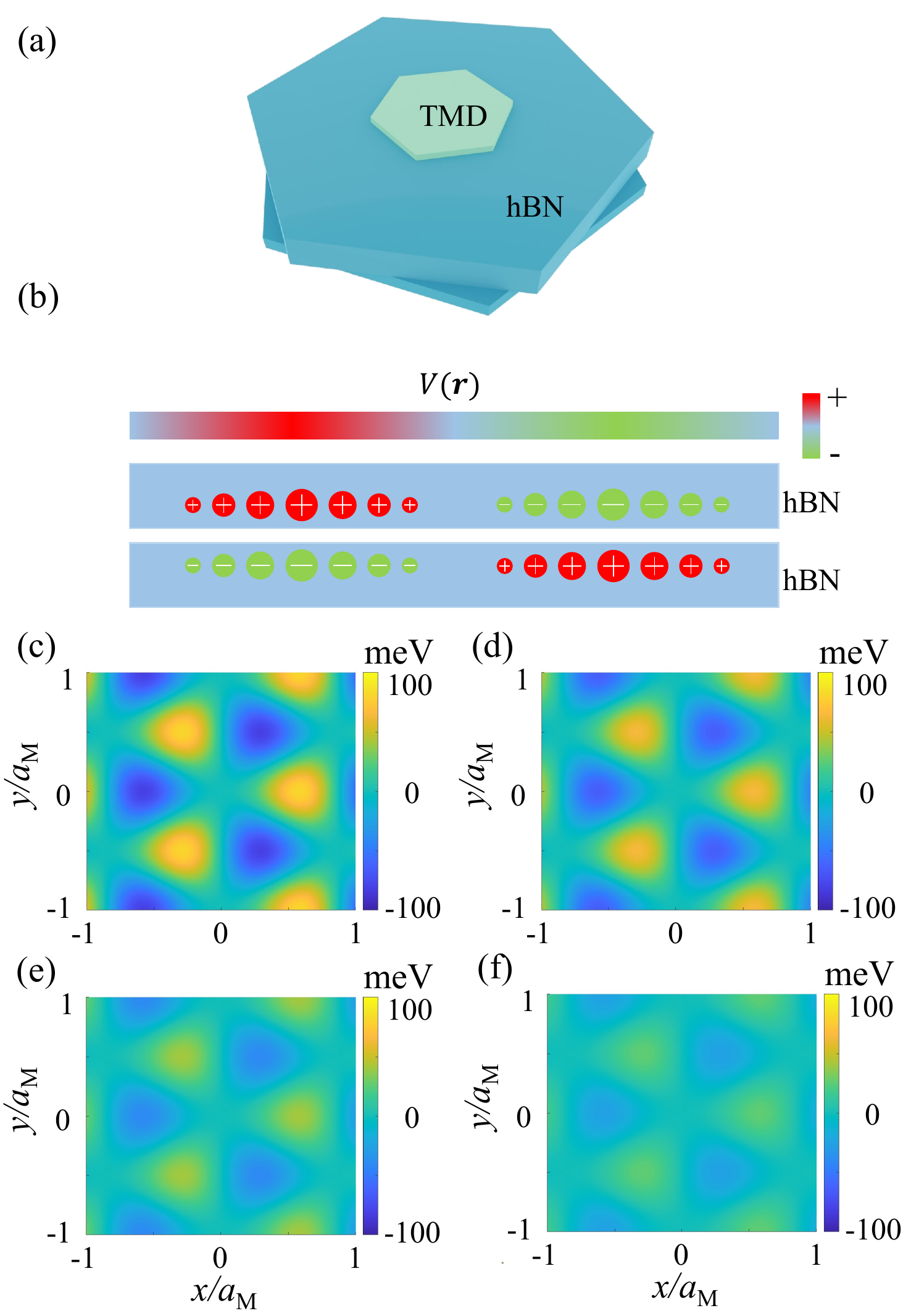}	
	\caption{(a) Schematic of the experimental setup: a TMD monolayer is placed atop a twisted hBN substrate, which itself is composed of multiple hBN layers stacked on another set of hBN multilayers with a controlled twist angle between them. The resulting moir\'e superlattice in the hBN substrate induces a periodic moir\'e potential for trions confined in the TMD monolayer. (b) At the interface between the twisted hBN multilayers, an electric polarization emerges, generating an electrostatic moir\'e potential $V(\boldsymbol{r})$ that affects the trions in the TMD layer. (c)-(f) Calculated spatial profile of the moir\'e potential when the capping hBN layer number is 1, 2, 4, and 6, respectively.} 
	\label{<fig_model>}%
\end{figure}

The key ingredient for realizing a topological trion insulator is to generate a moir\'e potential that acts on the trions. One straightforward way to achieve this is by placing a monolayer TMD on a substrate that provides a noninvasive moir\'e potential for the trions in the TMD.
 The substrate can be realized by the twisted multiplayer hBN, which provides a universal superlattice potential independent of the target layer \cite{zhaoUniversalSuperlatticePotential2021}, as illustrated in Fig. \ref{<fig_model>}(a). The twisted hBN substrate consists of $n$ layers of AA' stacked hBN on top of another $n$ layers of AA' stacked hBN. Spontaneous charge redistribution occurs at the twist interface, leading to electrical polarization and creating an electrostatic superlattice potential, $V_{\text{moir\'e}}$, that acts on the trions in the monolayer TMD.
 Remarkably, this structure has already been experimentally realized recently \cite{kimElectrostaticMoirePotential2024,wangMoireBandStructure2025,kiperConfinedTrionsMottWigner2025}. The moir\'e potential is tunable by the hBN layer number $n$ and twist angle $\theta$ (see Fig. \ref{<fig_model>}(b)). 
The moir\'e period is given by $ a_{M} = {a_{0}}/{\theta}$, where $a_{0}$ is the lattice constant of hBN, and $\theta$ is the twist angle between the hBN multilayers. The moir\'e potential energy is proportional to $\mathrm{exp}(-{4\pi} n d/{\sqrt{3}a_{M}})$, where $d$ is the thickness of a monolayer hBN. Given the moir\'e potential from the twisted hBN substrate, the Hamiltonian describing trions in the monolayer TMD is as follows \cite{zhangSinglePhotonEmitters2023}
\begin{equation}
	\begin{split}
		\hat{H}_{\text{trion }}&=\frac{\hbar^{2}\boldsymbol{Q}^{2}}{2m_{\text{trion}}}+\beta \boldsymbol{Q}+\beta \boldsymbol{Q}\cos(2\boldsymbol{\phi}_{\boldsymbol{\boldsymbol{Q}}})\sigma_{x}    \\
		&+\beta \boldsymbol{Q}\sin(2\boldsymbol{\phi}_{\boldsymbol{\boldsymbol{Q}}})\sigma_{y}+\frac{\delta}{2}(\sigma_{z}s_{z}+1)+V_{\text{moir\'e }}.
	\end{split}
 \label{h_trion}
\end{equation}
Here $m_{\text{trion}}\approx 1.6 m_{\text{e}}$ is the effective mass of the trion in TMD, and the exchange interaction strength is $\delta\approx6$ meV \cite{yuDiracConesDirac2014,hichriTrionFineStructure2020,courtadeChargedExcitonsMonolayer2017}.
The moir\'e potential created by the twisted hBN substrate can be expressed as $V_{\text{moir\'e }}=\Sigma_{j=1}^{6}V_{j}\exp(i\boldsymbol{G}_{j}\cdot \boldsymbol{r})$, where $\boldsymbol{G}_{j}$ is the moir\'e reciprocal lattice vector obtained by rotating $\boldsymbol{G}_{1}=(0, \frac{4\pi}{\sqrt{3}a_{M}})$ by an angle of $(j-1)\pi/3$.  Furthermore, $V_{j}=V\mathrm{exp}(-G n d)\mathrm{exp}(i\varphi\times(-1)^{j})$, where $G$ is the magnitude of the moir\'e reciprocal lattice vectors. First-principle calculations suggest that $V=-19.5$ meV and $\varphi=-\pi/2$ \cite{zhaoUniversalSuperlatticePotential2021}. In Figs. \ref{<fig_model>}(c)-\ref{<fig_model>}(f), we show the profile of the moir\'e periodic potential generated by the interface charge acting on the target material for hBN layer numbers of 1, 2, 4, and 6, respectively. Clearly, the strength of the moir\'e potential decreases monotonically as the number of hBN layers increases.

\begin{figure}
	\centering 
	\includegraphics[width=0.95\linewidth]{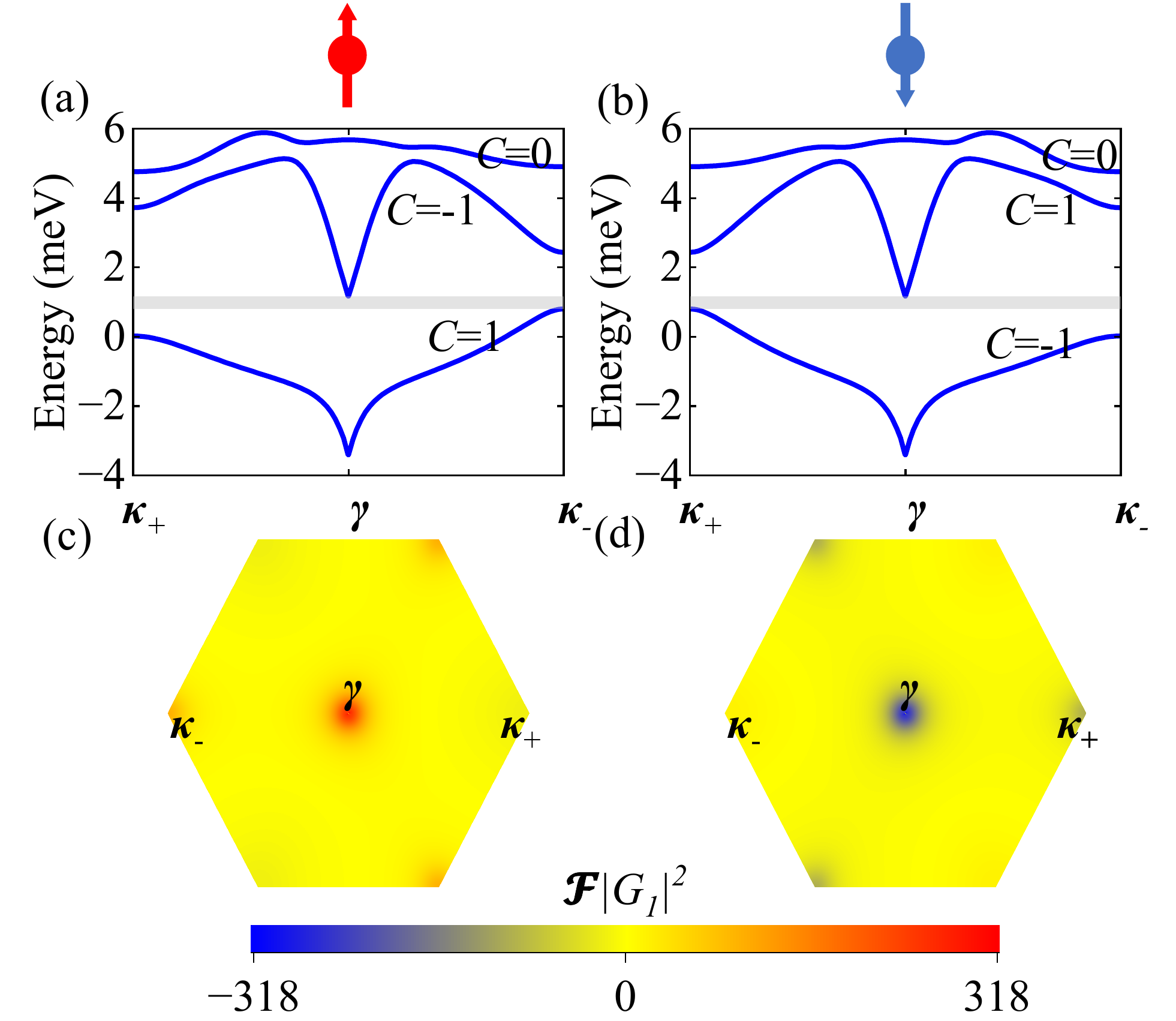}
	\caption{(a)-(b) The lowest three trion bands and their corresponding Chern numbers for the excess electron in the spin-up and spin-down states, respectively. (c)-(d) The Berry curvatures of the lowest trion band shown in panels (a) and (b), respectively. The calculations are performed at a twist angle $\theta = 1.1^\circ$ and $n = 10$.} 
	\label{<fig_energy_band>}%
\end{figure}

After diagonalizing the Hamiltonian, we obtain the trion band structure shown in Figs. \ref{<fig_energy_band>}(a)-\ref{<fig_energy_band>}(b) for different excess spin states, as well as the Chern numbers of the trion bands, calculated from the corresponding Berry curvature in Figs. \ref{<fig_energy_band>}(c)-\ref{<fig_energy_band>}(d). For the spin-up state, the ground state band has a Chern number $\mathit{C} = 1$, while the first excited state band has a Chern number $\mathit{C} = -1$. For the spin-down state, the Chern numbers for each band are opposite to those of the corresponding spin-up band. Since there is a global gap between the lowest and higher energy bands, the system exhibits nonzero and opposite Chern numbers for different spin states when the trion Fermi level—adjustable by both optical excitation intensity and gate voltage—lies within this band gap. As a result, the system has a nonzero $\mathbb{Z}_2$ number.

\begin{figure}[t]
	\centering 
	\includegraphics[width=0.95\linewidth]{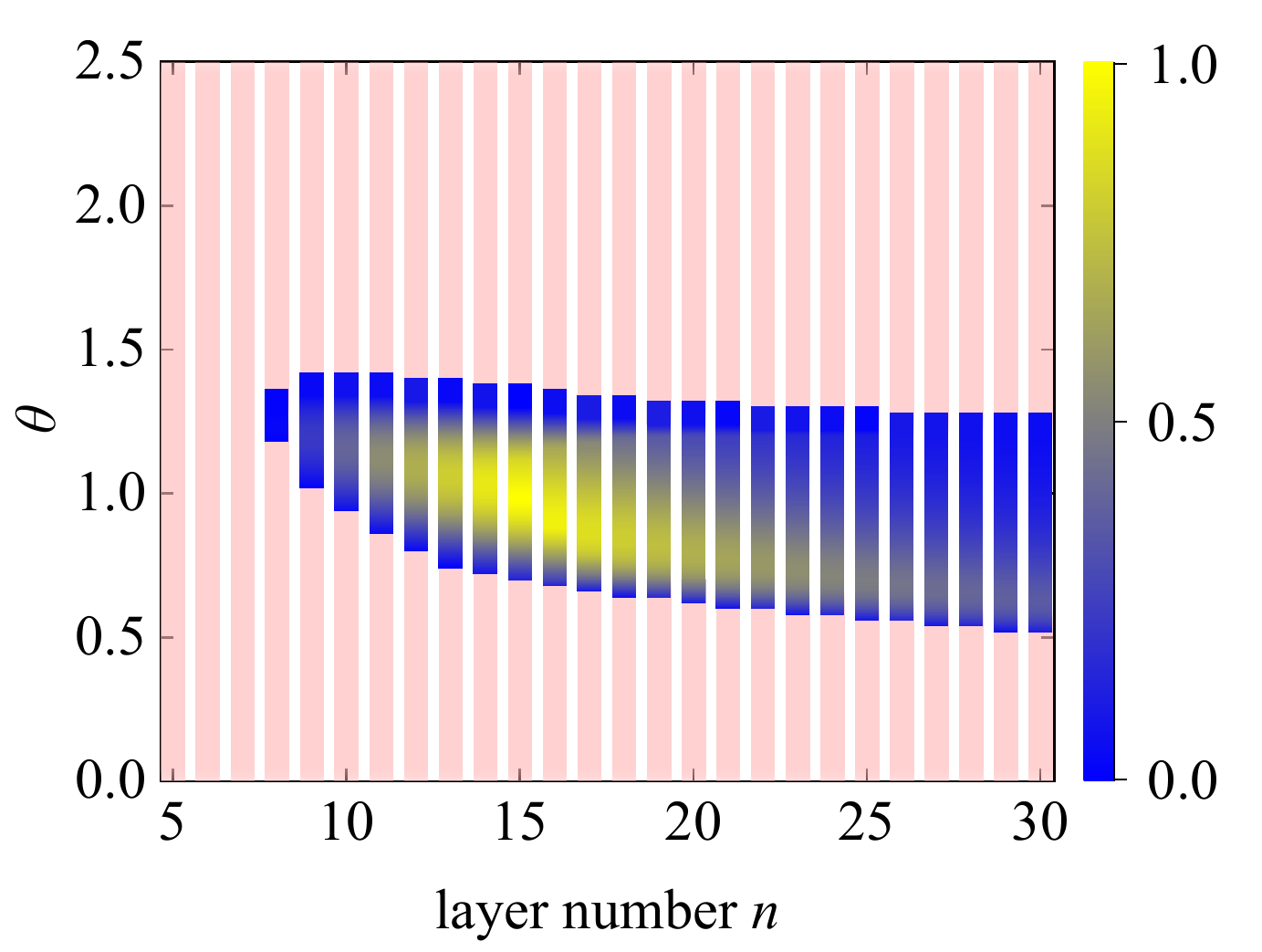}
	\caption{Phase diagram of the lowest trion bands in a TMD monolayer on the twisted hBN substrate, plotted as a function of hBN layer number $n$ and substrate twist angle $\theta$. The blue region indicates a $\mathbb{Z}_2$ topological trion insulator phase, with the color intensity representing the size of the energy gap. The red region corresponds to the topologically trivial phase.} 
	\label{<fig_phase>}%
\end{figure}

To systematically study the parameter regime for realizing topological trion bands, we calculate the $\mathbb{Z}_2$ number of the lowest trion band for different hBN layer numbers $n$ and twist angles $\theta$, and map out the phase diagram shown in Fig. \ref{<fig_phase>}. When the layer number $n\geq8$, there is a specific range of twist angles where the lowest trion band becomes topological. As the layer number increases, the range of twist angles supporting a topological trion band also broadens. However, with a large layer number, the moir\'e potential from the hBN substrate becomes weaker, leading to a much smaller topological band gap. Therefore, for easier experimental observation, one can choose a layer number $n\sim15$ and a twist angle $\theta\sim1^\circ$.

According to the bulk-edge correspondence, the $\mathbb{Z}_2$ topological trion insulator phase in the bulk should result in helical edge states, similar to those found in the $\mathbb{Z}_2$ topological insulator phase for electrons. To clearly demonstrate the presence of helical edge states for trions, we impose periodic boundary conditions in the $x$-direction and open boundary conditions in the $y$-direction. Mathematically, these boundary conditions are expressed as $\psi_{y=0}(k_{x}) = \psi_{y=L_y}(k_{x}) = 0$, where $k_x$ is the quasi-momentum along the $x$-direction and $y$ ranges from $0$ to $L_y$. We use the following wave function basis \cite{wuTopologicalExcitonBands2017}, which satisfies these boundary conditions, to construct the effective Hamiltonian under this mixed boundary setup, i.e.,
\begin{equation}
\left|{Q}_{{x}},m,\alpha\right\rangle =\sqrt{\frac{2}{\mathcal{A}}}\sin(\frac{m\pi}{L_{y}}y)\exp(i {Q}_{{x}} {x})|\alpha\rangle.
	\label{<eq_Substrate_transformation>}%
\end{equation}
Here $\mathcal{A}$ is the system area, $m$ labels the quantum number of the wave function in the $y$-direction and $|\alpha\rangle$ denotes the index of $\pm K$ valley exciton. Before expressing the Hamiltonian in the basis above, it is helpful to first rewrite the Hamiltonian in the following form
\begin{equation}
\hat{H}=\hat{h}_{0}+\hat{h}_{x}+\hat{h}_{y}+\hat{h}_{z}+\sum_{j=1}^{6}\tilde{V}_{j}\exp(i\boldsymbol{G}_{j}\cdot\boldsymbol{r})\sigma_{0},
	\label{<eq_part_Hamiltonian>}%
\end{equation}
where $\hat{h}_{0}=(\frac{\hbar^{2}\boldsymbol{Q}^{2}}{2m_{\text{trion}}}+\beta|\boldsymbol{Q}|)\sigma_{0}$, $\hat{h}_{x}=\beta|\boldsymbol{Q}|\cos(2\boldsymbol{\phi}_{\boldsymbol{Q}})\sigma_{x}$, $\hat{h}_{y}=\beta|\boldsymbol{Q}|\sin(2\boldsymbol{\phi}_{\boldsymbol{Q}})\sigma_{y}$,  $\hat{h}_{z}=\frac{\delta}{2}(\sigma_{z}s_{z}+1)$, and $\tilde{V}_{j}=V_{j}\exp(-i\boldsymbol{G}_{j}\cdot\boldsymbol{l})$.
In addition, $\boldsymbol{Q}$ represents the vector $(Q_{x}, \frac{m\pi}{L_{y}})$, and $\boldsymbol{l}$ is a vector which determines the spatial position of the edge states within the moir\'e unit cell \cite{wuTopologicalExcitonBands2017}.
In the basis given above, the matrix elements of the Hamiltonian are as follows,
\begin{widetext}
	\begin{align}
		\begin{split}
			\langle Q_x,m,\alpha|\hat{h}_0+\hat{h}_z|Q_x^\prime,m^\prime,\alpha^\prime\rangle&=\delta_{Q_x,Q_x^\prime}\delta_{mm^\prime}\delta_{\alpha\alpha^\prime}\left[\frac{\hbar^2\boldsymbol{Q}^2}{2m_{\text{trion}}}+\beta|\boldsymbol{Q}|+\frac{\delta}{2}(\sigma_{z}^{(\alpha\alpha^\prime)}s_{z}+1)\right],\\
			\langle Q_x,m,\alpha|\hat{h}_x|Q_x^{\prime},m^{\prime},\alpha^{\prime}\rangle&=\delta_{Q_x,Q_x^{\prime}}\delta_{mm^{\prime}}\beta|\boldsymbol{Q}|\cos(2\phi_{\boldsymbol{Q}})\sigma_x^{(\alpha\alpha^{\prime})},\\
			\langle Q_x,m,\alpha|\hat{h}_y|Q_x^{\prime},m^{\prime},\alpha^{\prime}\rangle&=\delta_{Q_x,Q_x^{\prime}}\beta\left(\frac{Q_x}{\sqrt{(L_yQ_x)^2+(m\pi)^2}}+\frac{Q_x}{\sqrt{(L_yQ_x)^2+(m^{\prime}\pi)^2}}\right)f(m,m^{\prime})\sigma_y^{(\alpha\alpha^{\prime})},\\
			\langle Q_x,m,\alpha|\exp(i\boldsymbol{G}\cdot\boldsymbol{r})\sigma_0|Q_x^{\prime},m^{\prime},\alpha^{\prime}\rangle=&\delta_{Q_x-Q_x^{\prime},G_x}\delta_{\alpha\alpha^{\prime}}\frac{1}{2}\bigg[F(G_yL_y/\pi+m-m^{\prime})+F(G_yL_y/\pi-m+m^{\prime})\\&-F(G_yL_y/\pi+m+m^{\prime})-F(G_yL_y/\pi-m-m^{\prime})\bigg],
		\end{split}
	\end{align}
\end{widetext}
where the two functions are
\begin{equation}
\begin{aligned}
      	f(m,m^{\prime})&=-i\pi m^{\prime}\frac{2}{L_{y}}\int_{0}^{L_{y}}\sin(\frac{m\pi}{L_{y}}y)\cos(\frac{m^{\prime}\pi}{L_{y}}y)dy\\&=\begin{cases}-i\frac{2mm^{\prime}}{m^2-m^{\prime2}}\left[1-(-1)^{m+m^{\prime}}\right],&m\neq m^{\prime}\\0,&m=m^{\prime}\end{cases},
\end{aligned}
\end{equation}
and
\begin{equation}
	\begin{aligned}
		F(z)=&\frac{1}{L_y}\int_0^{L_y}\exp(i\frac{z\pi}{L_y}y)dy\\=&\begin{cases}i\frac{1-(-1)^z}{z\pi},&z\neq0\\1,&z=0\end{cases}.
	\end{aligned}
\end{equation}

\begin{figure}[h]
	\centering 
	\includegraphics[width=0.95\linewidth]{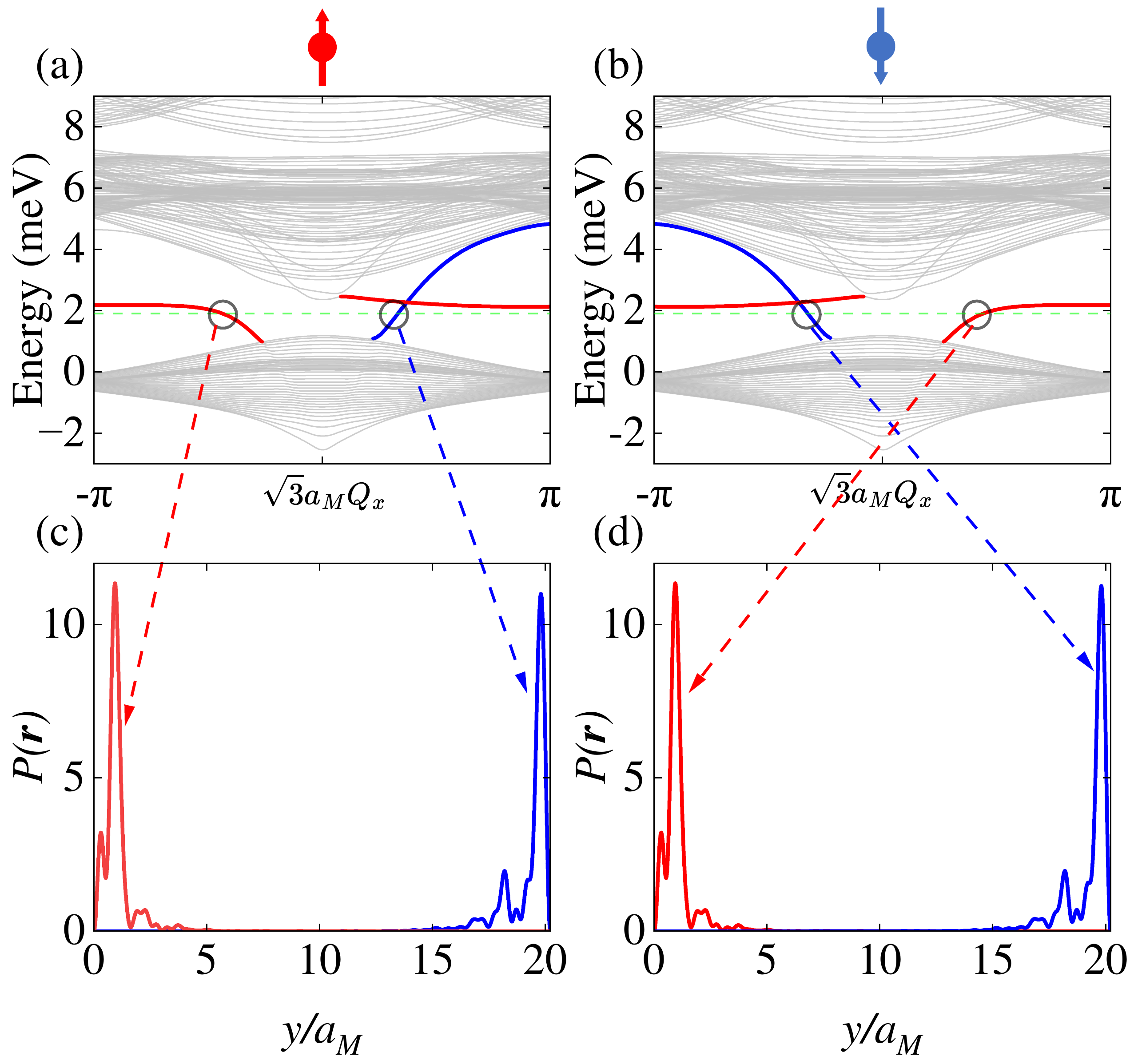}
	\caption{(a)-(b) Band structures of edge states (blue and red curves) and bulk states (gray curves) for excess electrons in spin-up and spin-down states, respectively. The calculations employ Eq. (\ref{<eq_part_Hamiltonian>}) with open boundary conditions along the $y$-direction and periodic boundary conditions along the $x$-direction, for $L_y = 20.2a_M$ and the vector $\boldsymbol{l} = (0,0.1)a_{M}$. The green dashed line denotes the Fermi level. The twist angle is $\theta = 1.1^\circ$ and $n=10$. (c)-(d) Spatial distributions of $P(\boldsymbol{r})$, as defined in Eq. (\ref{<eq_spatial distribution>}), for two edge states at the Fermi level shown in (a)-(b), plotted along the $y$-direction.} 
	\label{<fig_edge_state>}%
\end{figure}

The range of $Q_x$ is $(-\pi/\sqrt{3}a_{M}, \pi/\sqrt{3}a_{M})$, and $m$ is a positive integer. We choose an appropriate discretization for $Q_x$ and a truncation for $m$ to numerically diagonalize the Hamiltonian in the above basis. The resulting bands along the $Q_x$ direction are shown in Figs.  \ref{<fig_edge_state>}(a)-\ref{<fig_edge_state>}(b) for the spin-up and spin-down states, respectively. The red and blue dispersions connecting the conduction and valence bands represent gapless edge states with opposite group velocities. To further illustrate the spatial distribution of these edge states, we present the spatial profile of the following function
\begin{equation}
	P(\boldsymbol{r})=|\psi_{+K}(\boldsymbol{r})|^{2}+|\psi_{-K}(\boldsymbol{r})|^{2},
	\label{<eq_spatial distribution>}%
\end{equation}
where $(\psi_{+K}(\boldsymbol{r}),\psi_{-K}(\boldsymbol{r}))$ represents the edge state wave functions in $\pm K$ valley components, and $P(\boldsymbol{r})$ directly reveal the spatial distribution of edge states. We select the edge states at the Fermi level in Figs. \ref{<fig_edge_state>}(a)-\ref{<fig_edge_state>}(b), and plot the spatial profile $P(\boldsymbol{r})$ in Figs. \ref{<fig_edge_state>}(c)-\ref{<fig_edge_state>}(d). The spatial distribution of these edge states is clearly concentrated at the boundaries $y=0$ and $y=L_{y}$, confirming their edge state nature.
Based on the group velocities of these edge states, we find that each edge hosts a pair of oppositely propagating edge states, similar to the helical edge states in 2D quantum spin Hall insulators. The clear presence and spatial localization of these edge states confirm that our system is a $\mathbb{Z}_2$ topological insulator, exhibiting helical edge states of trions with dissipationless transport.

\begin{figure*}
	\centering 
	\includegraphics[width=0.95\linewidth]{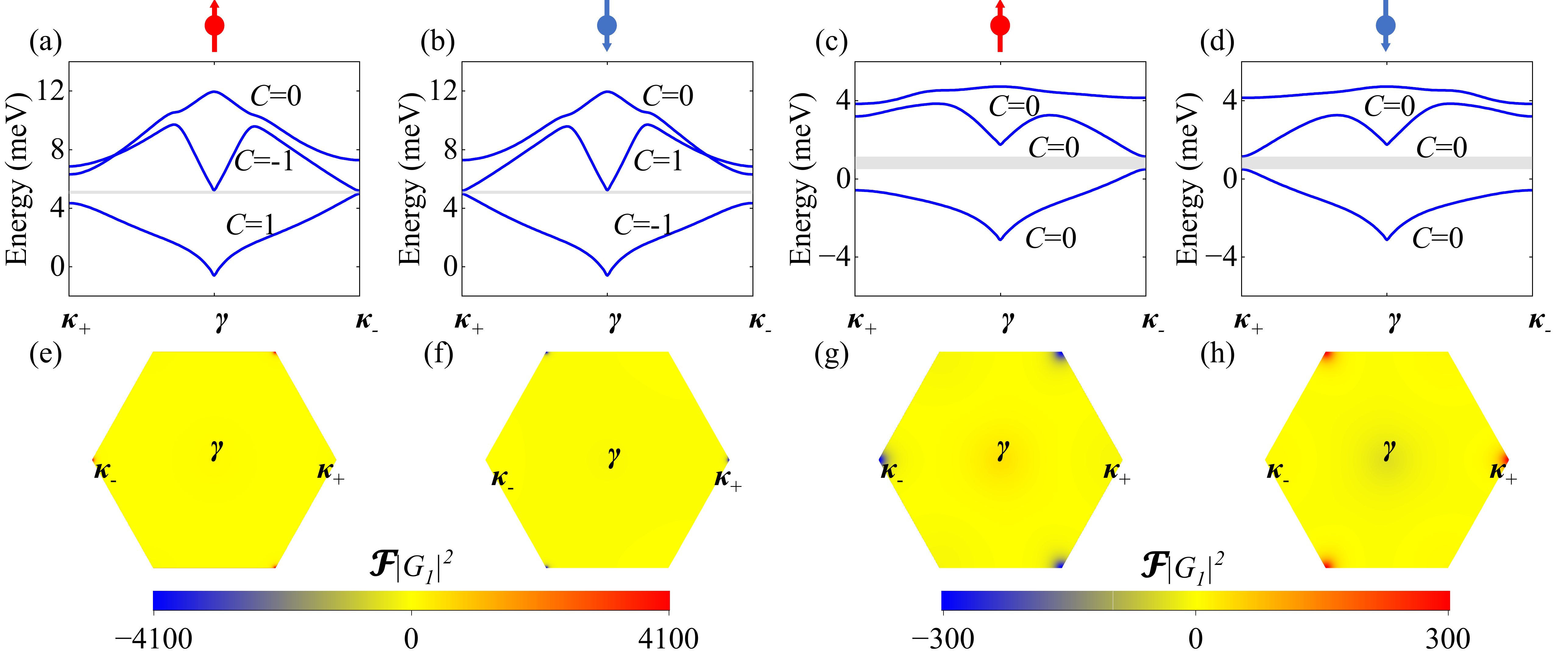}
	\caption{Trion band structures and Berry curvatures under screening effects (screening parameter $k_{s} = 0.15 \text{\AA}^{-1}$ and $\epsilon = (3.3 + 1)/2$, corresponding to uncapped TMD on hBN substrate). (a)-(b) Trion band structures for excess electrons in spin-up and spin-down states, respectively, at a twist angle $\theta = 1.3^\circ$ and $n = 13$. The lowest trion bands are topological. (c)-(d) Same as (a)-(b), but with twist angle $\theta = 1^\circ$ and $n = 13$. The lowest trion bands are topologically trivial. (e)-(h) Corresponding Berry curvatures for the lowest trion bands shown in panels (a)-(d).} 
	\label{<fig_screen>}%
\end{figure*}

In the calculations above, we have assumed the unscreened Coulomb potential for simplicity. However, in practical material systems, the screened Coulomb interaction should be considered based on the actual environment surrounding the TMD. In the following, we examine the stability of the $\mathbb{Z}_2$ topological number and helical edge states when screening is taken into account. We model the screening effect using the Rytova-Keldysh potential, as it provides very good agreement with experimental results for exciton Rydberg states in TMD \cite{stierMagnetoopticsExcitonRydberg2018,chernikovExcitonBindingEnergy2014}. In this case, the 2D screened Coulomb potential takes the form \cite{kylanpaaBindingEnergiesExciton2015,wangColloquiumExcitonsAtomically2018,hichriTrionFineStructure2020,courtadeChargedExcitonsMonolayer2017,stierMagnetoopticsExcitonRydberg2018}
\begin{equation}
	V(\rho) =-\frac{2\pi e^{2}}{\epsilon r_s}\left[\mathrm{H_{0}}\left(\frac{\rho}{r_{s}}\right)-Y_{0}\left(\frac{\rho}{r_{s}}\right)\right].
	\label{RKscreen>}%
\end{equation}
Here, $\rho$ is the in-plane distance between the electron and hole, $r_s$ is the effective screening radius, and  $\epsilon$ is the average dielectric constant of the substrate and capping layers, i.e., $\epsilon=(\epsilon_{\text{sub}}+\epsilon_{\text{cap}}
)/2$. $\mathrm{H_{0}}$ and $Y_0$ are the Struve and Neumann functions, respectively. We take $r_s=6.4\text{\AA}$ (with the screening parameter $k_{s}=1/r_s=0.15\text{\AA}^{-1}$), as this value gives a magnitude of exchange-induced splitting $\delta$ that agrees well with experimental results \cite{courtadeChargedExcitonsMonolayer2017}.
The Rytova-Keldysh potential in reciprocal space is given by \cite{hichriTrionFineStructure2020}
\begin{equation}
	V(Q) =-\frac{4\pi e^{2}}{\epsilon Q}\left(\frac{k_{s}}{k_{s}+Q}\right),
	\label{RKscreen_reciprocal>}%
\end{equation}
and accordingly, the strength of the electron-hole exchange interaction within the exciton can be obtained as 
\begin{equation}
	\beta_{\mathrm{scr}} =\beta\frac{k_{s}}{\epsilon(Q + k_{s})}.
	\label{screen>}%
\end{equation}
Thus, in the small $Q$ limit, screening reduces the exchange interaction strength by renormalizing $\beta$ to $\beta/\epsilon$. Using the screened electron-hole exchange interaction strength $\beta_{\mathrm{scr}}$, we recalculate the trion band structure and their topological properties, as shown in Figs. \ref{<fig_screen>}(a)-\ref{<fig_screen>}(b). For the twist angle $\theta = 1.3^{\circ}$ and $n = 13$, the lowest trion band remains topological, as shown in Figs.  \ref{<fig_screen>}(a)-\ref{<fig_screen>}(b), with the corresponding Berry curvature in Figs.  \ref{<fig_screen>}(e)-\ref{<fig_screen>}(f).  While for the twist angle $\theta = 1^{\circ}$ and $n = 13$, the lowest trion band becomes topologically trivial, as shown in Figs. \ref{<fig_screen>}(c)-\ref{<fig_screen>}(d), with the corresponding Berry curvature in Figs.  \ref{<fig_screen>}(g)-\ref{<fig_screen>}(h). Comparison with the phase diagram in Fig. \ref{<fig_phase>} shows that the screening effect can reduce the band gap between the lowest topological band and the first excited band, and can also turn a topological phase into a topologically trivial phase. In other words, the screening effect shrinks the region of the topological phase in the phase diagram.

\subsection{Twisted TMD heterbilayers}
\label{model2}
\begin{figure}[h]
	\centering 
	\includegraphics[width=0.95\linewidth]{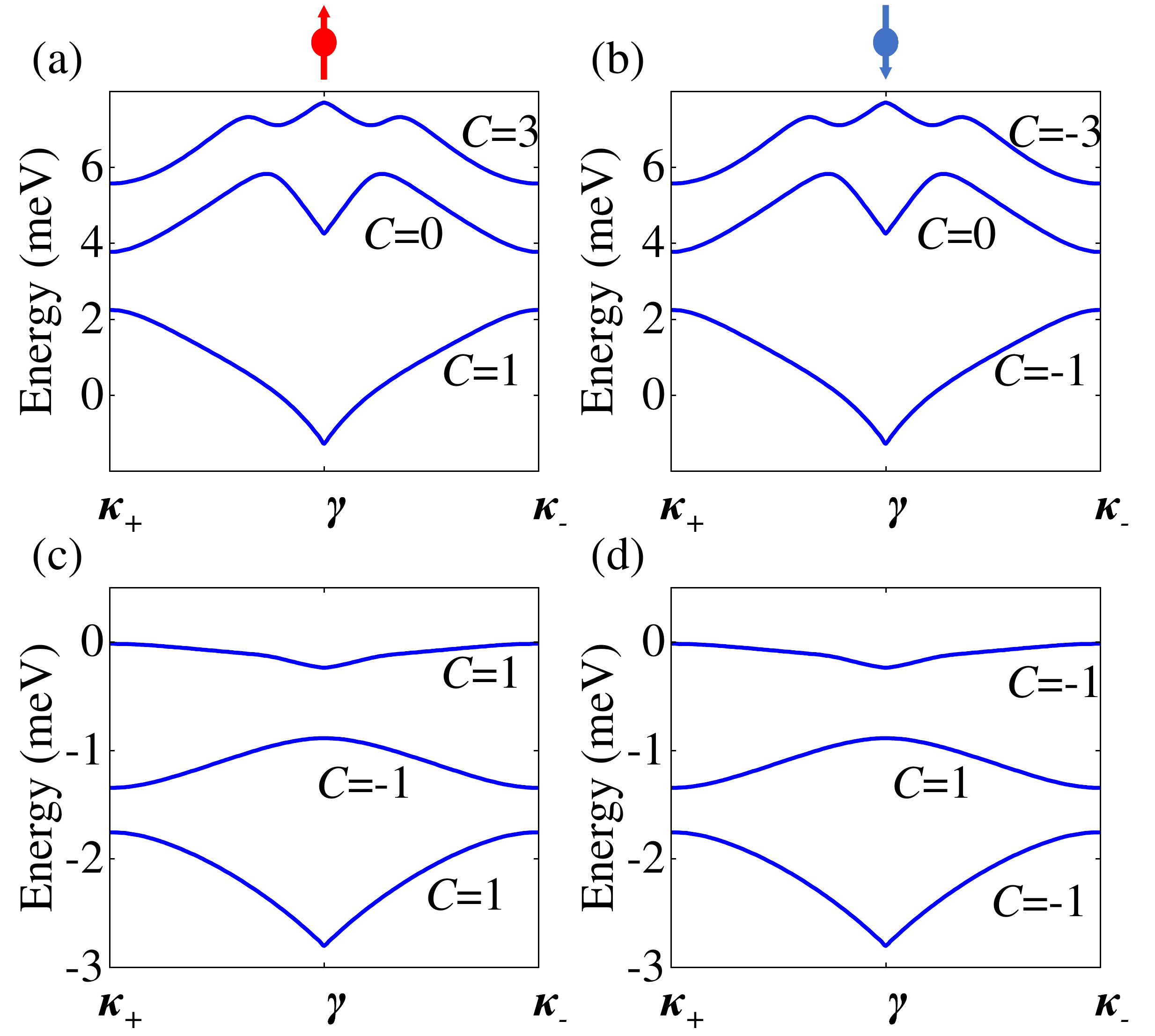}
	\caption{The lowest three trion bands and their corresponding Chern numbers for two different spin states in a twisted WSe$_2$/MoSe$_2$ heterobilayer. (a)-(b) Moir\'e period $a_M = 13$ nm and potential strength $V = 8$ meV. (c)-(d) Moir\'e period $a_M = 40$ nm and potential strength $V = 8$ meV. In all panels, the screening parameter is $k_{s} = 0.15$ $\text{\AA}^{-1}$ and the dielectric constant is $\epsilon = 3.3$, corresponding to TMD heterobilayers encapsulated with hBN.}
	\label{<fig_type_2>}%
\end{figure}

We now turn to another system which can realize the above proposed $\mathbb{Z}_2$ topological trion insulator. The system consist of twisted TMD heterobilayers, which have been extensively studied and the signature of moir\'e trions is observed \cite{liuSignaturesMoireTrions2021,wangMoireTrionsMoSe22021,brotons-gisbertMoireTrappedInterlayerTrions2021,marcellinaEvidenceMoireTrions2021}. In the harmonic approximation, each species of charged carriers experience the moir\'e potentials with the form \cite{liuSignaturesMoireTrions2021},
\begin{equation}
	V_{M}(\boldsymbol{r})=-\frac{V}{9}\sum_{j=1}^{6}\exp(i\boldsymbol{G}_{j}\cdot\boldsymbol{r})+\frac{V}{6}-\frac{|V|}{2},
	\label{<type_2_single>}%
\end{equation}
where $V$ is the moir\'e potential strength, and $\boldsymbol{G}_{j}$ represents the six moir\'e reciprocal lattice vectors, each with equal magnitude and rotated by $\pi/3$ from one another.
 For composite particles such as excitons or trions, the overall potential is the sum of the individual potentials of each constituent carrier, corrected by a form factor that accounts for the spatial extent of the composite particles. In particular, the moir\'e potential for trions in momentum space is given by
\begin{equation}
	V_{\mathrm{moir\acute{e}}}\left(\boldsymbol{g}-\boldsymbol{g}^{\prime}\right)=-\frac{V}{9}I_{0}\sum_{j=1}^{6}\delta_{\boldsymbol{G}_{j},\boldsymbol{g}-\boldsymbol{g}^{\prime}}-V\delta_{\boldsymbol{g},\boldsymbol{g}^{\prime}}.
	\label{<type_2_trion>}%
\end{equation}
Here, $I_{0}$ is the form factor accounting for the finite spatial extent of the trion wave function, which is determined by the specific profile of the trion’s internal state \cite{liuSignaturesMoireTrions2021}. $\boldsymbol{g}$ labels the trion plane wave basis, where hybridization occurs between states that differ by moir\'e reciprocal lattice vectors, leading to the formation of trion Bloch waves. 

We now use the twisted WSe$_2$/MoSe$_2$ heterobilayer as a specific example. A rich variety of composite particles, including intralayer and interlayer moir\'e excitons and trions, have been observed experimentally by resolving a series of sharp emission lines in the photoluminescence (PL) spectrum \cite{liuSignaturesMoireTrions2021}. In this system, the calculated form factor for the negatively charged trion is $I_{0} \approx 2.2$ \cite{liuSignaturesMoireTrions2021}. The moir\'e potential strength is set to $V = 8$ meV and is assumed to be independent of the moir\'e size within the parameter range of 13 nm to 40 nm \cite{liuSignaturesMoireTrions2021}. By plugging the above moir\'e potential for the trion into Eq. (\ref{h_trion}), we can readily obtain the intralayer moir\'e trion bands and the corresponding Chern numbers, similar to the approach in Sec. \ref{model1}. In Fig. \ref{<fig_type_2>}(a)-\ref{<fig_type_2>}(d), we present the lowest three moir\'e trion bands in the screened
case for different spin states, along with their corresponding Chern numbers. We find that the lowest trion band is topological for the moir\'e period $a_M=20$ nm (corresponding to the twist angle $\theta=0.9^\circ$), in the samples studied in the experiment \cite{liuSignaturesMoireTrions2021}. By varying the moir\'e period $a_M$ or the twist angle, we further find that within the range of $a_M = 13 \, \text{nm} \sim 40 \, \text{nm}$ (the twist angle within the range of $\theta=0.4^\circ \sim 1.4^\circ$) studied in the experiment \cite{liuSignaturesMoireTrions2021}, the lowest trion band remains topological and thus should host helical edge states. Therefore, this heterobilayer system can also serve as a platform for realizing the $\mathbb{Z}_2$ topological trion insulator.

\section{Discussion and conclusion}
\label{sec4}
The helical trion edge states can also serve as a dissipationless transport channel for electrons. So, in principle, these helical edge states can be detected through electrical transport measurements, similar to those used for quantum spin Hall insulators \cite{konigQuantumSpinHall2007,bernevigQuantumSpinHall2006}, within the trion lifetime ($\sim10$ ps). In addition, since the helical edge states have an exciton component that can radiatively recombine, they can also be detected using spatially resolved PL mapping, a technique previously used to observe the exciton Hall effect
\cite{ongaExcitonHallEffect2017,huangRobustRoomTemperature2020}. The helical edge state should appear as a spatially sharp peak in PL intensity localized at the sample boundary.
In twisted TMD heterobilayers, topological intralayer trions can be distinguished from other excitations such as excitons and interlayer trions by their distinct emission peaks in PL measurements.

By using ferromagnetic electrodes in contact with TMD, it is possible to achieve spin-selective injection of electrons \cite{liangElectricalSpinInjection2017}, allowing control over the trion spin species. In this way, one can realize a quantum anomalous Hall insulator of trions with a tunable Chern number and chiral edge states.

In summary, we have proposed the concept of a $\mathbb{Z}_2$ topological trion insulator in TMD monolayers and heterostructures. This state is achieved through the electron-hole exchange interaction within intralayer excitons and the exchange interaction between the excess electron and the intralayer excitons.
 This topological state of matter features helical edge states that act as dissipationless channels for trion transport, making it a promising building block for creating dissipationless exciton and trion circuits.
 We proposed two specific material implementations for the $\mathbb{Z}_2$ topological trion insulator and provided the phase diagram with respect to tunable parameters such as twist angle. We also confirmed that the topological phases remain stable under screening effects. This work could potentially support the development of TMD-based excitonic and trionic devices with dissipationless channels.

\begin{acknowledgements}
This work is supported by the National Key Research and Development Program of China (Grant No. 2022YFA1405304), the National Natural Science Foundation of China (Grant No. 12004118), and the Guangdong Provincial Quantum Science Strategic Initiative (Grant No. GDZX2401002).
\end{acknowledgements}


\begin{thebibliography}{70}%
\makeatletter
\providecommand \@ifxundefined [1]{%
 \@ifx{#1\undefined}
}%
\providecommand \@ifnum [1]{%
 \ifnum #1\expandafter \@firstoftwo
 \else \expandafter \@secondoftwo
 \fi
}%
\providecommand \@ifx [1]{%
 \ifx #1\expandafter \@firstoftwo
 \else \expandafter \@secondoftwo
 \fi
}%
\providecommand \natexlab [1]{#1}%
\providecommand \enquote  [1]{``#1''}%
\providecommand \bibnamefont  [1]{#1}%
\providecommand \bibfnamefont [1]{#1}%
\providecommand \citenamefont [1]{#1}%
\providecommand \href@noop [0]{\@secondoftwo}%
\providecommand \href [0]{\begingroup \@sanitize@url \@href}%
\providecommand \@href[1]{\@@startlink{#1}\@@href}%
\providecommand \@@href[1]{\endgroup#1\@@endlink}%
\providecommand \@sanitize@url [0]{\catcode `\\12\catcode `\$12\catcode
  `\&12\catcode `\#12\catcode `\^12\catcode `\_12\catcode `\%12\relax}%
\providecommand \@@startlink[1]{}%
\providecommand \@@endlink[0]{}%
\providecommand \url  [0]{\begingroup\@sanitize@url \@url }%
\providecommand \@url [1]{\endgroup\@href {#1}{\urlprefix }}%
\providecommand \urlprefix  [0]{URL }%
\providecommand \Eprint [0]{\href }%
\providecommand \doibase [0]{https://doi.org/}%
\providecommand \selectlanguage [0]{\@gobble}%
\providecommand \bibinfo  [0]{\@secondoftwo}%
\providecommand \bibfield  [0]{\@secondoftwo}%
\providecommand \translation [1]{[#1]}%
\providecommand \BibitemOpen [0]{}%
\providecommand \bibitemStop [0]{}%
\providecommand \bibitemNoStop [0]{.\EOS\space}%
\providecommand \EOS [0]{\spacefactor3000\relax}%
\providecommand \BibitemShut  [1]{\csname bibitem#1\endcsname}%
\let\auto@bib@innerbib\@empty
\bibitem [{\citenamefont {He}\ \emph {et~al.}(2014)\citenamefont {He},
  \citenamefont {Kumar}, \citenamefont {Zhao}, \citenamefont {Wang},
  \citenamefont {Mak}, \citenamefont {Zhao},\ and\ \citenamefont
  {Shan}}]{heTightlyBoundExcitons2014}%
  \BibitemOpen
  \bibfield  {author} {\bibinfo {author} {\bibfnamefont {K.}~\bibnamefont
  {He}}, \bibinfo {author} {\bibfnamefont {N.}~\bibnamefont {Kumar}}, \bibinfo
  {author} {\bibfnamefont {L.}~\bibnamefont {Zhao}}, \bibinfo {author}
  {\bibfnamefont {Z.}~\bibnamefont {Wang}}, \bibinfo {author} {\bibfnamefont
  {K.~F.}\ \bibnamefont {Mak}}, \bibinfo {author} {\bibfnamefont
  {H.}~\bibnamefont {Zhao}},\ and\ \bibinfo {author} {\bibfnamefont
  {J.}~\bibnamefont {Shan}},\ }\href
  {https://link.aps.org/doi/10.1103/PhysRevLett.113.026803} {\bibfield
  {journal} {\bibinfo  {journal} {Phys. Rev. Lett.}\ }\textbf {\bibinfo
  {volume} {113}},\ \bibinfo {pages} {026803} (\bibinfo {year}
  {2014})}\BibitemShut {NoStop}%
\bibitem [{\citenamefont {Sallen}\ \emph {et~al.}(2012)\citenamefont {Sallen},
  \citenamefont {Bouet}, \citenamefont {Marie}, \citenamefont {Wang},
  \citenamefont {Zhu}, \citenamefont {Han}, \citenamefont {Lu}, \citenamefont
  {Tan}, \citenamefont {Amand}, \citenamefont {Liu},\ and\ \citenamefont
  {Urbaszek}}]{sallenRobustOpticalEmission2012}%
  \BibitemOpen
  \bibfield  {author} {\bibinfo {author} {\bibfnamefont {G.}~\bibnamefont
  {Sallen}}, \bibinfo {author} {\bibfnamefont {L.}~\bibnamefont {Bouet}},
  \bibinfo {author} {\bibfnamefont {X.}~\bibnamefont {Marie}}, \bibinfo
  {author} {\bibfnamefont {G.}~\bibnamefont {Wang}}, \bibinfo {author}
  {\bibfnamefont {C.~R.}\ \bibnamefont {Zhu}}, \bibinfo {author} {\bibfnamefont
  {W.~P.}\ \bibnamefont {Han}}, \bibinfo {author} {\bibfnamefont
  {Y.}~\bibnamefont {Lu}}, \bibinfo {author} {\bibfnamefont {P.~H.}\
  \bibnamefont {Tan}}, \bibinfo {author} {\bibfnamefont {T.}~\bibnamefont
  {Amand}}, \bibinfo {author} {\bibfnamefont {B.~L.}\ \bibnamefont {Liu}},\
  and\ \bibinfo {author} {\bibfnamefont {B.}~\bibnamefont {Urbaszek}},\ }\href
  {https://link.aps.org/doi/10.1103/PhysRevB.86.081301} {\bibfield  {journal}
  {\bibinfo  {journal} {Phys. Rev. B}\ }\textbf {\bibinfo {volume} {86}},\
  \bibinfo {pages} {081301} (\bibinfo {year} {2012})}\BibitemShut {NoStop}%
\bibitem [{\citenamefont {Korn}\ \emph {et~al.}(2011)\citenamefont {Korn},
  \citenamefont {Heydrich}, \citenamefont {Hirmer}, \citenamefont
  {Schmutzler},\ and\ \citenamefont
  {Sch{\"u}ller}}]{kornLowtemperaturePhotocarrierDynamics2011}%
  \BibitemOpen
  \bibfield  {author} {\bibinfo {author} {\bibfnamefont {T.}~\bibnamefont
  {Korn}}, \bibinfo {author} {\bibfnamefont {S.}~\bibnamefont {Heydrich}},
  \bibinfo {author} {\bibfnamefont {M.}~\bibnamefont {Hirmer}}, \bibinfo
  {author} {\bibfnamefont {J.}~\bibnamefont {Schmutzler}},\ and\ \bibinfo
  {author} {\bibfnamefont {C.}~\bibnamefont {Sch{\"u}ller}},\ }\href
  {https://pubs.aip.org/aip/apl/article/121914} {\bibfield  {journal} {\bibinfo
   {journal} {Appl. Phys. Lett.}\ }\textbf {\bibinfo {volume} {99}},\ \bibinfo
  {pages} {102109} (\bibinfo {year} {2011})}\BibitemShut {NoStop}%
\bibitem [{\citenamefont {Ugeda}\ \emph {et~al.}(2014)\citenamefont {Ugeda},
  \citenamefont {Bradley}, \citenamefont {Shi}, \citenamefont {Da~Jornada},
  \citenamefont {Zhang}, \citenamefont {Qiu}, \citenamefont {Ruan},
  \citenamefont {Mo}, \citenamefont {Hussain}, \citenamefont {Shen},
  \citenamefont {Wang}, \citenamefont {Louie},\ and\ \citenamefont
  {Crommie}}]{ugedaGiantBandgapRenormalization2014}%
  \BibitemOpen
  \bibfield  {author} {\bibinfo {author} {\bibfnamefont {M.~M.}\ \bibnamefont
  {Ugeda}}, \bibinfo {author} {\bibfnamefont {A.~J.}\ \bibnamefont {Bradley}},
  \bibinfo {author} {\bibfnamefont {S.-F.}\ \bibnamefont {Shi}}, \bibinfo
  {author} {\bibfnamefont {F.~H.}\ \bibnamefont {Da~Jornada}}, \bibinfo
  {author} {\bibfnamefont {Y.}~\bibnamefont {Zhang}}, \bibinfo {author}
  {\bibfnamefont {D.~Y.}\ \bibnamefont {Qiu}}, \bibinfo {author} {\bibfnamefont
  {W.}~\bibnamefont {Ruan}}, \bibinfo {author} {\bibfnamefont {S.-K.}\
  \bibnamefont {Mo}}, \bibinfo {author} {\bibfnamefont {Z.}~\bibnamefont
  {Hussain}}, \bibinfo {author} {\bibfnamefont {Z.-X.}\ \bibnamefont {Shen}},
  \bibinfo {author} {\bibfnamefont {F.}~\bibnamefont {Wang}}, \bibinfo {author}
  {\bibfnamefont {S.~G.}\ \bibnamefont {Louie}},\ and\ \bibinfo {author}
  {\bibfnamefont {M.~F.}\ \bibnamefont {Crommie}},\ }\href
  {https://www.nature.com/articles/nmat4061} {\bibfield  {journal} {\bibinfo
  {journal} {Nat. Mater.}\ }\textbf {\bibinfo {volume} {13}},\ \bibinfo {pages}
  {1091} (\bibinfo {year} {2014})}\BibitemShut {NoStop}%
\bibitem [{\citenamefont {Wang}\ \emph {et~al.}(2015)\citenamefont {Wang},
  \citenamefont {Marie}, \citenamefont {Gerber}, \citenamefont {Amand},
  \citenamefont {Lagarde}, \citenamefont {Bouet}, \citenamefont {Vidal},
  \citenamefont {Balocchi},\ and\ \citenamefont
  {Urbaszek}}]{wangGiantEnhancementOptical2015}%
  \BibitemOpen
  \bibfield  {author} {\bibinfo {author} {\bibfnamefont {G.}~\bibnamefont
  {Wang}}, \bibinfo {author} {\bibfnamefont {X.}~\bibnamefont {Marie}},
  \bibinfo {author} {\bibfnamefont {I.}~\bibnamefont {Gerber}}, \bibinfo
  {author} {\bibfnamefont {T.}~\bibnamefont {Amand}}, \bibinfo {author}
  {\bibfnamefont {D.}~\bibnamefont {Lagarde}}, \bibinfo {author} {\bibfnamefont
  {L.}~\bibnamefont {Bouet}}, \bibinfo {author} {\bibfnamefont
  {M.}~\bibnamefont {Vidal}}, \bibinfo {author} {\bibfnamefont
  {A.}~\bibnamefont {Balocchi}},\ and\ \bibinfo {author} {\bibfnamefont
  {B.}~\bibnamefont {Urbaszek}},\ }\href
  {https://link.aps.org/doi/10.1103/PhysRevLett.114.097403} {\bibfield
  {journal} {\bibinfo  {journal} {Phys. Rev. Lett.}\ }\textbf {\bibinfo
  {volume} {114}},\ \bibinfo {pages} {097403} (\bibinfo {year}
  {2015})}\BibitemShut {NoStop}%
\bibitem [{\citenamefont {Chernikov}\ \emph {et~al.}(2014)\citenamefont
  {Chernikov}, \citenamefont {Berkelbach}, \citenamefont {Hill}, \citenamefont
  {Rigosi}, \citenamefont {Li}, \citenamefont {Aslan}, \citenamefont
  {Reichman}, \citenamefont {Hybertsen},\ and\ \citenamefont
  {Heinz}}]{chernikovExcitonBindingEnergy2014}%
  \BibitemOpen
  \bibfield  {author} {\bibinfo {author} {\bibfnamefont {A.}~\bibnamefont
  {Chernikov}}, \bibinfo {author} {\bibfnamefont {T.~C.}\ \bibnamefont
  {Berkelbach}}, \bibinfo {author} {\bibfnamefont {H.~M.}\ \bibnamefont
  {Hill}}, \bibinfo {author} {\bibfnamefont {A.}~\bibnamefont {Rigosi}},
  \bibinfo {author} {\bibfnamefont {Y.}~\bibnamefont {Li}}, \bibinfo {author}
  {\bibfnamefont {B.}~\bibnamefont {Aslan}}, \bibinfo {author} {\bibfnamefont
  {D.~R.}\ \bibnamefont {Reichman}}, \bibinfo {author} {\bibfnamefont {M.~S.}\
  \bibnamefont {Hybertsen}},\ and\ \bibinfo {author} {\bibfnamefont {T.~F.}\
  \bibnamefont {Heinz}},\ }\href
  {https://link.aps.org/doi/10.1103/PhysRevLett.113.076802} {\bibfield
  {journal} {\bibinfo  {journal} {Phys. Rev. Lett.}\ }\textbf {\bibinfo
  {volume} {113}},\ \bibinfo {pages} {076802} (\bibinfo {year}
  {2014})}\BibitemShut {NoStop}%
\bibitem [{\citenamefont {Qiu}\ \emph {et~al.}(2013)\citenamefont {Qiu},
  \citenamefont {Da~Jornada},\ and\ \citenamefont
  {Louie}}]{qiuOpticalSpectrumOfMoS22013}%
  \BibitemOpen
  \bibfield  {author} {\bibinfo {author} {\bibfnamefont {D.~Y.}\ \bibnamefont
  {Qiu}}, \bibinfo {author} {\bibfnamefont {F.~H.}\ \bibnamefont
  {Da~Jornada}},\ and\ \bibinfo {author} {\bibfnamefont {S.~G.}\ \bibnamefont
  {Louie}},\ }\href {https://link.aps.org/doi/10.1103/PhysRevLett.111.216805}
  {\bibfield  {journal} {\bibinfo  {journal} {Phys. Rev. Lett.}\ }\textbf
  {\bibinfo {volume} {111}},\ \bibinfo {pages} {216805} (\bibinfo {year}
  {2013})}\BibitemShut {NoStop}%
\bibitem [{\citenamefont {Komsa}\ and\ \citenamefont
  {Krasheninnikov}(2012)}]{komsaEffectsConfinementEnvironment2012}%
  \BibitemOpen
  \bibfield  {author} {\bibinfo {author} {\bibfnamefont {H.-P.}\ \bibnamefont
  {Komsa}}\ and\ \bibinfo {author} {\bibfnamefont {A.~V.}\ \bibnamefont
  {Krasheninnikov}},\ }\href
  {https://link.aps.org/doi/10.1103/PhysRevB.86.241201} {\bibfield  {journal}
  {\bibinfo  {journal} {Phys. Rev. B}\ }\textbf {\bibinfo {volume} {86}},\
  \bibinfo {pages} {241201} (\bibinfo {year} {2012})}\BibitemShut {NoStop}%
\bibitem [{\citenamefont {Hanbicki}\ \emph {et~al.}(2015)\citenamefont
  {Hanbicki}, \citenamefont {Currie}, \citenamefont {Kioseoglou}, \citenamefont
  {Friedman},\ and\ \citenamefont
  {Jonker}}]{hanbickiMeasurementHighExciton2015}%
  \BibitemOpen
  \bibfield  {author} {\bibinfo {author} {\bibfnamefont {A.}~\bibnamefont
  {Hanbicki}}, \bibinfo {author} {\bibfnamefont {M.}~\bibnamefont {Currie}},
  \bibinfo {author} {\bibfnamefont {G.}~\bibnamefont {Kioseoglou}}, \bibinfo
  {author} {\bibfnamefont {A.}~\bibnamefont {Friedman}},\ and\ \bibinfo
  {author} {\bibfnamefont {B.}~\bibnamefont {Jonker}},\ }\href
  {https://linkinghub.elsevier.com/retrieve/pii/S0038109814004621} {\bibfield
  {journal} {\bibinfo  {journal} {Solid State Commun.}\ }\textbf {\bibinfo
  {volume} {203}},\ \bibinfo {pages} {16} (\bibinfo {year} {2015})}\BibitemShut
  {NoStop}%
\bibitem [{\citenamefont {Hsu}\ \emph {et~al.}(2019)\citenamefont {Hsu},
  \citenamefont {Quan}, \citenamefont {Wang}, \citenamefont {Lu}, \citenamefont
  {Campbell}, \citenamefont {Chang}, \citenamefont {Li}, \citenamefont {Li},\
  and\ \citenamefont {Shih}}]{hsuDielectricImpactExciton2019}%
  \BibitemOpen
  \bibfield  {author} {\bibinfo {author} {\bibfnamefont {W.-T.}\ \bibnamefont
  {Hsu}}, \bibinfo {author} {\bibfnamefont {J.}~\bibnamefont {Quan}}, \bibinfo
  {author} {\bibfnamefont {C.-Y.}\ \bibnamefont {Wang}}, \bibinfo {author}
  {\bibfnamefont {L.-S.}\ \bibnamefont {Lu}}, \bibinfo {author} {\bibfnamefont
  {M.}~\bibnamefont {Campbell}}, \bibinfo {author} {\bibfnamefont {W.-H.}\
  \bibnamefont {Chang}}, \bibinfo {author} {\bibfnamefont {L.-J.}\ \bibnamefont
  {Li}}, \bibinfo {author} {\bibfnamefont {X.}~\bibnamefont {Li}},\ and\
  \bibinfo {author} {\bibfnamefont {C.-K.}\ \bibnamefont {Shih}},\ }\href
  {https://iopscience.iop.org/article/10.1088/2053-1583/ab072a} {\bibfield
  {journal} {\bibinfo  {journal} {2D Mater.}\ }\textbf {\bibinfo {volume}
  {6}},\ \bibinfo {pages} {025028} (\bibinfo {year} {2019})}\BibitemShut
  {NoStop}%
\bibitem [{\citenamefont {Park}\ \emph {et~al.}(2018)\citenamefont {Park},
  \citenamefont {Mutz}, \citenamefont {Schultz}, \citenamefont {Blumstengel},
  \citenamefont {Han}, \citenamefont {Aljarb}, \citenamefont {Li},
  \citenamefont {{List-Kratochvil}}, \citenamefont {Amsalem},\ and\
  \citenamefont {Koch}}]{parkDirectDeterminationMonolayer2018}%
  \BibitemOpen
  \bibfield  {author} {\bibinfo {author} {\bibfnamefont {S.}~\bibnamefont
  {Park}}, \bibinfo {author} {\bibfnamefont {N.}~\bibnamefont {Mutz}}, \bibinfo
  {author} {\bibfnamefont {T.}~\bibnamefont {Schultz}}, \bibinfo {author}
  {\bibfnamefont {S.}~\bibnamefont {Blumstengel}}, \bibinfo {author}
  {\bibfnamefont {A.}~\bibnamefont {Han}}, \bibinfo {author} {\bibfnamefont
  {A.}~\bibnamefont {Aljarb}}, \bibinfo {author} {\bibfnamefont {L.-J.}\
  \bibnamefont {Li}}, \bibinfo {author} {\bibfnamefont {E.~J.~W.}\ \bibnamefont
  {{List-Kratochvil}}}, \bibinfo {author} {\bibfnamefont {P.}~\bibnamefont
  {Amsalem}},\ and\ \bibinfo {author} {\bibfnamefont {N.}~\bibnamefont
  {Koch}},\ }\href
  {https://iopscience.iop.org/article/10.1088/2053-1583/aaa4ca} {\bibfield
  {journal} {\bibinfo  {journal} {2D Mater.}\ }\textbf {\bibinfo {volume}
  {5}},\ \bibinfo {pages} {025003} (\bibinfo {year} {2018})}\BibitemShut
  {NoStop}%
\bibitem [{\citenamefont {Kyl{\"a}np{\"a}{\"a}}\ and\ \citenamefont
  {Komsa}(2015)}]{kylanpaaBindingEnergiesExciton2015}%
  \BibitemOpen
  \bibfield  {author} {\bibinfo {author} {\bibfnamefont {I.}~\bibnamefont
  {Kyl{\"a}np{\"a}{\"a}}}\ and\ \bibinfo {author} {\bibfnamefont {H.-P.}\
  \bibnamefont {Komsa}},\ }\href
  {https://link.aps.org/doi/10.1103/PhysRevB.92.205418} {\bibfield  {journal}
  {\bibinfo  {journal} {Phys. Rev. B}\ }\textbf {\bibinfo {volume} {92}},\
  \bibinfo {pages} {205418} (\bibinfo {year} {2015})}\BibitemShut {NoStop}%
\bibitem [{\citenamefont {Zhu}\ \emph {et~al.}(2015)\citenamefont {Zhu},
  \citenamefont {Chen},\ and\ \citenamefont
  {Cui}}]{zhuExcitonBindingEnergy2015}%
  \BibitemOpen
  \bibfield  {author} {\bibinfo {author} {\bibfnamefont {B.}~\bibnamefont
  {Zhu}}, \bibinfo {author} {\bibfnamefont {X.}~\bibnamefont {Chen}},\ and\
  \bibinfo {author} {\bibfnamefont {X.}~\bibnamefont {Cui}},\ }\href
  {https://www.nature.com/articles/srep09218} {\bibfield  {journal} {\bibinfo
  {journal} {Sci. Rep.}\ }\textbf {\bibinfo {volume} {5}},\ \bibinfo {pages}
  {09218} (\bibinfo {year} {2015})}\BibitemShut {NoStop}%
\bibitem [{\citenamefont {Yu}\ \emph {et~al.}(2015{\natexlab{a}})\citenamefont
  {Yu}, \citenamefont {Wang}, \citenamefont {Tong}, \citenamefont {Xu},\ and\
  \citenamefont {Yao}}]{yuAnomalousLightCones2015}%
  \BibitemOpen
  \bibfield  {author} {\bibinfo {author} {\bibfnamefont {H.}~\bibnamefont
  {Yu}}, \bibinfo {author} {\bibfnamefont {Y.}~\bibnamefont {Wang}}, \bibinfo
  {author} {\bibfnamefont {Q.}~\bibnamefont {Tong}}, \bibinfo {author}
  {\bibfnamefont {X.}~\bibnamefont {Xu}},\ and\ \bibinfo {author}
  {\bibfnamefont {W.}~\bibnamefont {Yao}},\ }\href
  {https://link.aps.org/doi/10.1103/PhysRevLett.115.187002} {\bibfield
  {journal} {\bibinfo  {journal} {Phys. Rev. Lett.}\ }\textbf {\bibinfo
  {volume} {115}},\ \bibinfo {pages} {187002} (\bibinfo {year}
  {2015}{\natexlab{a}})}\BibitemShut {NoStop}%
\bibitem [{\citenamefont {Liu}\ \emph {et~al.}(2014)\citenamefont {Liu},
  \citenamefont {Shen}, \citenamefont {Su}, \citenamefont {Hsu}, \citenamefont
  {Li},\ and\ \citenamefont {Li}}]{liuOpticalPropertiesMonolayer2014}%
  \BibitemOpen
  \bibfield  {author} {\bibinfo {author} {\bibfnamefont {H.-L.}\ \bibnamefont
  {Liu}}, \bibinfo {author} {\bibfnamefont {C.-C.}\ \bibnamefont {Shen}},
  \bibinfo {author} {\bibfnamefont {S.-H.}\ \bibnamefont {Su}}, \bibinfo
  {author} {\bibfnamefont {C.-L.}\ \bibnamefont {Hsu}}, \bibinfo {author}
  {\bibfnamefont {M.-Y.}\ \bibnamefont {Li}},\ and\ \bibinfo {author}
  {\bibfnamefont {L.-J.}\ \bibnamefont {Li}},\ }\href
  {https://pubs.aip.org/apl/article/105/20/201905/27433/Optical-properties-of-monolayer-transition-metal}
  {\bibfield  {journal} {\bibinfo  {journal} {Appl. Phys. Lett.}\ }\textbf
  {\bibinfo {volume} {105}},\ \bibinfo {pages} {201905} (\bibinfo {year}
  {2014})}\BibitemShut {NoStop}%
\bibitem [{\citenamefont {Ye}\ \emph {et~al.}(2014)\citenamefont {Ye},
  \citenamefont {Cao}, \citenamefont {O'Brien}, \citenamefont {Zhu},
  \citenamefont {Yin}, \citenamefont {Wang}, \citenamefont {Louie},\ and\
  \citenamefont {Zhang}}]{yeProbingExcitonicDark2014}%
  \BibitemOpen
  \bibfield  {author} {\bibinfo {author} {\bibfnamefont {Z.}~\bibnamefont
  {Ye}}, \bibinfo {author} {\bibfnamefont {T.}~\bibnamefont {Cao}}, \bibinfo
  {author} {\bibfnamefont {K.}~\bibnamefont {O'Brien}}, \bibinfo {author}
  {\bibfnamefont {H.}~\bibnamefont {Zhu}}, \bibinfo {author} {\bibfnamefont
  {X.}~\bibnamefont {Yin}}, \bibinfo {author} {\bibfnamefont {Y.}~\bibnamefont
  {Wang}}, \bibinfo {author} {\bibfnamefont {S.~G.}\ \bibnamefont {Louie}},\
  and\ \bibinfo {author} {\bibfnamefont {X.}~\bibnamefont {Zhang}},\ }\href
  {https://www.nature.com/articles/nature13734} {\bibfield  {journal} {\bibinfo
   {journal} {Nature}\ }\textbf {\bibinfo {volume} {513}},\ \bibinfo {pages}
  {214} (\bibinfo {year} {2014})}\BibitemShut {NoStop}%
\bibitem [{\citenamefont {Xiao}\ \emph {et~al.}(2015)\citenamefont {Xiao},
  \citenamefont {Ye}, \citenamefont {Wang}, \citenamefont {Zhu}, \citenamefont
  {Wang},\ and\ \citenamefont {Zhang}}]{xiaoNonlinearOpticalSelection2015}%
  \BibitemOpen
  \bibfield  {author} {\bibinfo {author} {\bibfnamefont {J.}~\bibnamefont
  {Xiao}}, \bibinfo {author} {\bibfnamefont {Z.}~\bibnamefont {Ye}}, \bibinfo
  {author} {\bibfnamefont {Y.}~\bibnamefont {Wang}}, \bibinfo {author}
  {\bibfnamefont {H.}~\bibnamefont {Zhu}}, \bibinfo {author} {\bibfnamefont
  {Y.}~\bibnamefont {Wang}},\ and\ \bibinfo {author} {\bibfnamefont
  {X.}~\bibnamefont {Zhang}},\ }\href
  {https://www.nature.com/articles/lsa2015139} {\bibfield  {journal} {\bibinfo
  {journal} {Light Sci. Appl.}\ }\textbf {\bibinfo {volume} {4}},\ \bibinfo
  {pages} {e366} (\bibinfo {year} {2015})}\BibitemShut {NoStop}%
\bibitem [{\citenamefont {Courtade}\ \emph {et~al.}(2017)\citenamefont
  {Courtade}, \citenamefont {Semina}, \citenamefont {Manca}, \citenamefont
  {Glazov}, \citenamefont {Robert}, \citenamefont {Cadiz}, \citenamefont
  {Wang}, \citenamefont {Taniguchi}, \citenamefont {Watanabe}, \citenamefont
  {Pierre}, \citenamefont {Escoffier}, \citenamefont {Ivchenko}, \citenamefont
  {Renucci}, \citenamefont {Marie}, \citenamefont {Amand},\ and\ \citenamefont
  {Urbaszek}}]{courtadeChargedExcitonsMonolayer2017}%
  \BibitemOpen
  \bibfield  {author} {\bibinfo {author} {\bibfnamefont {E.}~\bibnamefont
  {Courtade}}, \bibinfo {author} {\bibfnamefont {M.}~\bibnamefont {Semina}},
  \bibinfo {author} {\bibfnamefont {M.}~\bibnamefont {Manca}}, \bibinfo
  {author} {\bibfnamefont {M.~M.}\ \bibnamefont {Glazov}}, \bibinfo {author}
  {\bibfnamefont {C.}~\bibnamefont {Robert}}, \bibinfo {author} {\bibfnamefont
  {F.}~\bibnamefont {Cadiz}}, \bibinfo {author} {\bibfnamefont
  {G.}~\bibnamefont {Wang}}, \bibinfo {author} {\bibfnamefont {T.}~\bibnamefont
  {Taniguchi}}, \bibinfo {author} {\bibfnamefont {K.}~\bibnamefont {Watanabe}},
  \bibinfo {author} {\bibfnamefont {M.}~\bibnamefont {Pierre}}, \bibinfo
  {author} {\bibfnamefont {W.}~\bibnamefont {Escoffier}}, \bibinfo {author}
  {\bibfnamefont {E.~L.}\ \bibnamefont {Ivchenko}}, \bibinfo {author}
  {\bibfnamefont {P.}~\bibnamefont {Renucci}}, \bibinfo {author} {\bibfnamefont
  {X.}~\bibnamefont {Marie}}, \bibinfo {author} {\bibfnamefont
  {T.}~\bibnamefont {Amand}},\ and\ \bibinfo {author} {\bibfnamefont
  {B.}~\bibnamefont {Urbaszek}},\ }\href
  {http://link.aps.org/doi/10.1103/PhysRevB.96.085302} {\bibfield  {journal}
  {\bibinfo  {journal} {Phys. Rev. B}\ }\textbf {\bibinfo {volume} {96}},\
  \bibinfo {pages} {085302} (\bibinfo {year} {2017})}\BibitemShut {NoStop}%
\bibitem [{\citenamefont {Vaclavkova}\ \emph {et~al.}(2018)\citenamefont
  {Vaclavkova}, \citenamefont {Wyzula}, \citenamefont {Nogajewski},
  \citenamefont {Bartos}, \citenamefont {Slobodeniuk}, \citenamefont
  {Faugeras}, \citenamefont {Potemski},\ and\ \citenamefont
  {Molas}}]{vaclavkovaSingletTripletTrions2018}%
  \BibitemOpen
  \bibfield  {author} {\bibinfo {author} {\bibfnamefont {D.}~\bibnamefont
  {Vaclavkova}}, \bibinfo {author} {\bibfnamefont {J.}~\bibnamefont {Wyzula}},
  \bibinfo {author} {\bibfnamefont {K.}~\bibnamefont {Nogajewski}}, \bibinfo
  {author} {\bibfnamefont {M.}~\bibnamefont {Bartos}}, \bibinfo {author}
  {\bibfnamefont {A.~O.}\ \bibnamefont {Slobodeniuk}}, \bibinfo {author}
  {\bibfnamefont {C.}~\bibnamefont {Faugeras}}, \bibinfo {author}
  {\bibfnamefont {M.}~\bibnamefont {Potemski}},\ and\ \bibinfo {author}
  {\bibfnamefont {M.~R.}\ \bibnamefont {Molas}},\ }\href
  {https://iopscience.iop.org/article/10.1088/1361-6528/aac65c} {\bibfield
  {journal} {\bibinfo  {journal} {Nanotechnology}\ }\textbf {\bibinfo {volume}
  {29}},\ \bibinfo {pages} {325705} (\bibinfo {year} {2018})}\BibitemShut
  {NoStop}%
\bibitem [{\citenamefont {Jadczak}\ \emph {et~al.}(2017)\citenamefont
  {Jadczak}, \citenamefont {{Kutrowska-Girzycka}}, \citenamefont
  {Kapu{\'s}ci{\'n}ski}, \citenamefont {Huang}, \citenamefont {W{\'o}js},\ and\
  \citenamefont {Bryja}}]{jadczakProbingFreeLocalized2017}%
  \BibitemOpen
  \bibfield  {author} {\bibinfo {author} {\bibfnamefont {J.}~\bibnamefont
  {Jadczak}}, \bibinfo {author} {\bibfnamefont {J.}~\bibnamefont
  {{Kutrowska-Girzycka}}}, \bibinfo {author} {\bibfnamefont {P.}~\bibnamefont
  {Kapu{\'s}ci{\'n}ski}}, \bibinfo {author} {\bibfnamefont {Y.~S.}\
  \bibnamefont {Huang}}, \bibinfo {author} {\bibfnamefont {A.}~\bibnamefont
  {W{\'o}js}},\ and\ \bibinfo {author} {\bibfnamefont {L.}~\bibnamefont
  {Bryja}},\ }\href
  {https://iopscience.iop.org/article/10.1088/1361-6528/aa87d0} {\bibfield
  {journal} {\bibinfo  {journal} {Nanotechnology}\ }\textbf {\bibinfo {volume}
  {28}},\ \bibinfo {pages} {395702} (\bibinfo {year} {2017})}\BibitemShut
  {NoStop}%
\bibitem [{\citenamefont {Jakubczyk}\ \emph {et~al.}(2018)\citenamefont
  {Jakubczyk}, \citenamefont {Nogajewski}, \citenamefont {Molas}, \citenamefont
  {Bartos}, \citenamefont {Langbein}, \citenamefont {Potemski},\ and\
  \citenamefont {Kasprzak}}]{jakubczykImpactEnvironmentDynamics2018}%
  \BibitemOpen
  \bibfield  {author} {\bibinfo {author} {\bibfnamefont {T.}~\bibnamefont
  {Jakubczyk}}, \bibinfo {author} {\bibfnamefont {K.}~\bibnamefont
  {Nogajewski}}, \bibinfo {author} {\bibfnamefont {M.~R.}\ \bibnamefont
  {Molas}}, \bibinfo {author} {\bibfnamefont {M.}~\bibnamefont {Bartos}},
  \bibinfo {author} {\bibfnamefont {W.}~\bibnamefont {Langbein}}, \bibinfo
  {author} {\bibfnamefont {M.}~\bibnamefont {Potemski}},\ and\ \bibinfo
  {author} {\bibfnamefont {J.}~\bibnamefont {Kasprzak}},\ }\href
  {https://iopscience.iop.org/article/10.1088/2053-1583/aabc1c} {\bibfield
  {journal} {\bibinfo  {journal} {2D Mater.}\ }\textbf {\bibinfo {volume}
  {5}},\ \bibinfo {pages} {031007} (\bibinfo {year} {2018})}\BibitemShut
  {NoStop}%
\bibitem [{\citenamefont {Szyniszewski}\ \emph {et~al.}(2017)\citenamefont
  {Szyniszewski}, \citenamefont {Mostaani}, \citenamefont {Drummond},\ and\
  \citenamefont {Fal'ko}}]{szyniszewskiBindingEnergiesTrions2017}%
  \BibitemOpen
  \bibfield  {author} {\bibinfo {author} {\bibfnamefont {M.}~\bibnamefont
  {Szyniszewski}}, \bibinfo {author} {\bibfnamefont {E.}~\bibnamefont
  {Mostaani}}, \bibinfo {author} {\bibfnamefont {N.~D.}\ \bibnamefont
  {Drummond}},\ and\ \bibinfo {author} {\bibfnamefont {V.~I.}\ \bibnamefont
  {Fal'ko}},\ }\href {https://link.aps.org/doi/10.1103/PhysRevB.95.081301}
  {\bibfield  {journal} {\bibinfo  {journal} {Phys. Rev. B}\ }\textbf {\bibinfo
  {volume} {95}},\ \bibinfo {pages} {081301} (\bibinfo {year}
  {2017})}\BibitemShut {NoStop}%
\bibitem [{\citenamefont {Van Der~Donck}\ \emph {et~al.}(2017)\citenamefont
  {Van Der~Donck}, \citenamefont {Zarenia},\ and\ \citenamefont
  {Peeters}}]{vanderdonckExcitonsTrionsMonolayer2017}%
  \BibitemOpen
  \bibfield  {author} {\bibinfo {author} {\bibfnamefont {M.}~\bibnamefont {Van
  Der~Donck}}, \bibinfo {author} {\bibfnamefont {M.}~\bibnamefont {Zarenia}},\
  and\ \bibinfo {author} {\bibfnamefont {F.~M.}\ \bibnamefont {Peeters}},\
  }\href {http://link.aps.org/doi/10.1103/PhysRevB.96.035131} {\bibfield
  {journal} {\bibinfo  {journal} {Phys. Rev. B}\ }\textbf {\bibinfo {volume}
  {96}},\ \bibinfo {pages} {035131} (\bibinfo {year} {2017})}\BibitemShut
  {NoStop}%
\bibitem [{\citenamefont {Berkelbach}\ \emph {et~al.}(2013)\citenamefont
  {Berkelbach}, \citenamefont {Hybertsen},\ and\ \citenamefont
  {Reichman}}]{berkelbachTheoryNeutralCharged2013}%
  \BibitemOpen
  \bibfield  {author} {\bibinfo {author} {\bibfnamefont {T.~C.}\ \bibnamefont
  {Berkelbach}}, \bibinfo {author} {\bibfnamefont {M.~S.}\ \bibnamefont
  {Hybertsen}},\ and\ \bibinfo {author} {\bibfnamefont {D.~R.}\ \bibnamefont
  {Reichman}},\ }\href {https://link.aps.org/doi/10.1103/PhysRevB.88.045318}
  {\bibfield  {journal} {\bibinfo  {journal} {Phys. Rev. B}\ }\textbf {\bibinfo
  {volume} {88}},\ \bibinfo {pages} {045318} (\bibinfo {year}
  {2013})}\BibitemShut {NoStop}%
\bibitem [{\citenamefont {Mak}\ \emph {et~al.}(2013)\citenamefont {Mak},
  \citenamefont {He}, \citenamefont {Lee}, \citenamefont {Lee}, \citenamefont
  {Hone}, \citenamefont {Heinz},\ and\ \citenamefont
  {Shan}}]{makTightlyBoundTrions2013}%
  \BibitemOpen
  \bibfield  {author} {\bibinfo {author} {\bibfnamefont {K.~F.}\ \bibnamefont
  {Mak}}, \bibinfo {author} {\bibfnamefont {K.}~\bibnamefont {He}}, \bibinfo
  {author} {\bibfnamefont {C.}~\bibnamefont {Lee}}, \bibinfo {author}
  {\bibfnamefont {G.~H.}\ \bibnamefont {Lee}}, \bibinfo {author} {\bibfnamefont
  {J.}~\bibnamefont {Hone}}, \bibinfo {author} {\bibfnamefont {T.~F.}\
  \bibnamefont {Heinz}},\ and\ \bibinfo {author} {\bibfnamefont
  {J.}~\bibnamefont {Shan}},\ }\href {https://www.nature.com/articles/nmat3505}
  {\bibfield  {journal} {\bibinfo  {journal} {Nat. Mater.}\ }\textbf {\bibinfo
  {volume} {12}},\ \bibinfo {pages} {207} (\bibinfo {year} {2013})}\BibitemShut
  {NoStop}%
\bibitem [{\citenamefont {Lui}\ \emph {et~al.}(2014)\citenamefont {Lui},
  \citenamefont {Frenzel}, \citenamefont {Pilon}, \citenamefont {Lee},
  \citenamefont {Ling}, \citenamefont {Akselrod}, \citenamefont {Kong},\ and\
  \citenamefont {Gedik}}]{luiTrionInducedNegativePhotoconductivity2014}%
  \BibitemOpen
  \bibfield  {author} {\bibinfo {author} {\bibfnamefont {C.~H.}\ \bibnamefont
  {Lui}}, \bibinfo {author} {\bibfnamefont {A.~J.}\ \bibnamefont {Frenzel}},
  \bibinfo {author} {\bibfnamefont {D.~V.}\ \bibnamefont {Pilon}}, \bibinfo
  {author} {\bibfnamefont {Y.-H.}\ \bibnamefont {Lee}}, \bibinfo {author}
  {\bibfnamefont {X.}~\bibnamefont {Ling}}, \bibinfo {author} {\bibfnamefont
  {G.~M.}\ \bibnamefont {Akselrod}}, \bibinfo {author} {\bibfnamefont
  {J.}~\bibnamefont {Kong}},\ and\ \bibinfo {author} {\bibfnamefont
  {N.}~\bibnamefont {Gedik}},\ }\href
  {https://link.aps.org/doi/10.1103/PhysRevLett.113.166801} {\bibfield
  {journal} {\bibinfo  {journal} {Phys. Rev. Lett.}\ }\textbf {\bibinfo
  {volume} {113}},\ \bibinfo {pages} {166801} (\bibinfo {year}
  {2014})}\BibitemShut {NoStop}%
\bibitem [{\citenamefont {Filinov}\ \emph {et~al.}(2004)\citenamefont
  {Filinov}, \citenamefont {Riva}, \citenamefont {Peeters}, \citenamefont
  {Lozovik},\ and\ \citenamefont
  {Bonitz}}]{filinovInfluenceWellwidthFluctuations2004}%
  \BibitemOpen
  \bibfield  {author} {\bibinfo {author} {\bibfnamefont {A.~V.}\ \bibnamefont
  {Filinov}}, \bibinfo {author} {\bibfnamefont {C.}~\bibnamefont {Riva}},
  \bibinfo {author} {\bibfnamefont {F.~M.}\ \bibnamefont {Peeters}}, \bibinfo
  {author} {\bibfnamefont {{\relax Yu}.~E.}\ \bibnamefont {Lozovik}},\ and\
  \bibinfo {author} {\bibfnamefont {M.}~\bibnamefont {Bonitz}},\ }\href
  {https://link.aps.org/doi/10.1103/PhysRevB.70.035323} {\bibfield  {journal}
  {\bibinfo  {journal} {Phys. Rev. B}\ }\textbf {\bibinfo {volume} {70}},\
  \bibinfo {pages} {035323} (\bibinfo {year} {2004})}\BibitemShut {NoStop}%
\bibitem [{\citenamefont {Yu}\ \emph {et~al.}(2014)\citenamefont {Yu},
  \citenamefont {Liu}, \citenamefont {Gong}, \citenamefont {Xu},\ and\
  \citenamefont {Yao}}]{yuDiracConesDirac2014}%
  \BibitemOpen
  \bibfield  {author} {\bibinfo {author} {\bibfnamefont {H.}~\bibnamefont
  {Yu}}, \bibinfo {author} {\bibfnamefont {G.-B.}\ \bibnamefont {Liu}},
  \bibinfo {author} {\bibfnamefont {P.}~\bibnamefont {Gong}}, \bibinfo {author}
  {\bibfnamefont {X.}~\bibnamefont {Xu}},\ and\ \bibinfo {author}
  {\bibfnamefont {W.}~\bibnamefont {Yao}},\ }\href
  {https://www.nature.com/articles/ncomms4876} {\bibfield  {journal} {\bibinfo
  {journal} {Nat. Commun.}\ }\textbf {\bibinfo {volume} {5}},\ \bibinfo {pages}
  {3876} (\bibinfo {year} {2014})}\BibitemShut {NoStop}%
\bibitem [{\citenamefont {Hichri}\ and\ \citenamefont
  {Jaziri}(2020)}]{hichriTrionFineStructure2020}%
  \BibitemOpen
  \bibfield  {author} {\bibinfo {author} {\bibfnamefont {A.}~\bibnamefont
  {Hichri}}\ and\ \bibinfo {author} {\bibfnamefont {S.}~\bibnamefont
  {Jaziri}},\ }\href {https://link.aps.org/doi/10.1103/PhysRevB.102.085407}
  {\bibfield  {journal} {\bibinfo  {journal} {Phys. Rev. B}\ }\textbf {\bibinfo
  {volume} {102}},\ \bibinfo {pages} {085407} (\bibinfo {year}
  {2020})}\BibitemShut {NoStop}%
\bibitem [{\citenamefont {{Perea-Causin}}\ \emph {et~al.}(2022)\citenamefont
  {{Perea-Causin}}, \citenamefont {Brem},\ and\ \citenamefont
  {Malic}}]{perea-causinTrionphononInteractionAtomically2022}%
  \BibitemOpen
  \bibfield  {author} {\bibinfo {author} {\bibfnamefont {R.}~\bibnamefont
  {{Perea-Causin}}}, \bibinfo {author} {\bibfnamefont {S.}~\bibnamefont
  {Brem}},\ and\ \bibinfo {author} {\bibfnamefont {E.}~\bibnamefont {Malic}},\
  }\href {https://link.aps.org/doi/10.1103/PhysRevB.106.115407} {\bibfield
  {journal} {\bibinfo  {journal} {Phys. Rev. B}\ }\textbf {\bibinfo {volume}
  {106}},\ \bibinfo {pages} {115407} (\bibinfo {year} {2022})}\BibitemShut
  {NoStop}%
\bibitem [{\citenamefont {Meier}\ \emph {et~al.}(2023)\citenamefont {Meier},
  \citenamefont {Zhumagulov}, \citenamefont {Dietl}, \citenamefont {Parzefall},
  \citenamefont {Kempf}, \citenamefont {Holler}, \citenamefont {Nagler},
  \citenamefont {Faria~Junior}, \citenamefont {Fabian}, \citenamefont {Korn},\
  and\ \citenamefont {Sch{\"u}ller}}]{meierEmergentTrionphononCoupling2023}%
  \BibitemOpen
  \bibfield  {author} {\bibinfo {author} {\bibfnamefont {S.}~\bibnamefont
  {Meier}}, \bibinfo {author} {\bibfnamefont {Y.}~\bibnamefont {Zhumagulov}},
  \bibinfo {author} {\bibfnamefont {M.}~\bibnamefont {Dietl}}, \bibinfo
  {author} {\bibfnamefont {P.}~\bibnamefont {Parzefall}}, \bibinfo {author}
  {\bibfnamefont {M.}~\bibnamefont {Kempf}}, \bibinfo {author} {\bibfnamefont
  {J.}~\bibnamefont {Holler}}, \bibinfo {author} {\bibfnamefont
  {P.}~\bibnamefont {Nagler}}, \bibinfo {author} {\bibfnamefont {P.~E.}\
  \bibnamefont {Faria~Junior}}, \bibinfo {author} {\bibfnamefont
  {J.}~\bibnamefont {Fabian}}, \bibinfo {author} {\bibfnamefont
  {T.}~\bibnamefont {Korn}},\ and\ \bibinfo {author} {\bibfnamefont
  {C.}~\bibnamefont {Sch{\"u}ller}},\ }\href
  {https://link.aps.org/doi/10.1103/PhysRevResearch.5.L032036} {\bibfield
  {journal} {\bibinfo  {journal} {Phys. Rev. Res.}\ }\textbf {\bibinfo {volume}
  {5}},\ \bibinfo {pages} {L032036} (\bibinfo {year} {2023})}\BibitemShut
  {NoStop}%
\bibitem [{\citenamefont {Singh}\ \emph
  {et~al.}(2016{\natexlab{a}})\citenamefont {Singh}, \citenamefont {Moody},
  \citenamefont {Tran}, \citenamefont {Scott}, \citenamefont {Overbeck},
  \citenamefont {Bergh{\"a}user}, \citenamefont {Schaibley}, \citenamefont
  {Seifert}, \citenamefont {Pleskot}, \citenamefont {Gabor}, \citenamefont
  {Yan}, \citenamefont {Mandrus}, \citenamefont {Richter}, \citenamefont
  {Malic}, \citenamefont {Xu},\ and\ \citenamefont
  {Li}}]{singhTrionFormationDynamics2016}%
  \BibitemOpen
  \bibfield  {author} {\bibinfo {author} {\bibfnamefont {A.}~\bibnamefont
  {Singh}}, \bibinfo {author} {\bibfnamefont {G.}~\bibnamefont {Moody}},
  \bibinfo {author} {\bibfnamefont {K.}~\bibnamefont {Tran}}, \bibinfo {author}
  {\bibfnamefont {M.~E.}\ \bibnamefont {Scott}}, \bibinfo {author}
  {\bibfnamefont {V.}~\bibnamefont {Overbeck}}, \bibinfo {author}
  {\bibfnamefont {G.}~\bibnamefont {Bergh{\"a}user}}, \bibinfo {author}
  {\bibfnamefont {J.}~\bibnamefont {Schaibley}}, \bibinfo {author}
  {\bibfnamefont {E.~J.}\ \bibnamefont {Seifert}}, \bibinfo {author}
  {\bibfnamefont {D.}~\bibnamefont {Pleskot}}, \bibinfo {author} {\bibfnamefont
  {N.~M.}\ \bibnamefont {Gabor}}, \bibinfo {author} {\bibfnamefont
  {J.}~\bibnamefont {Yan}}, \bibinfo {author} {\bibfnamefont {D.~G.}\
  \bibnamefont {Mandrus}}, \bibinfo {author} {\bibfnamefont {M.}~\bibnamefont
  {Richter}}, \bibinfo {author} {\bibfnamefont {E.}~\bibnamefont {Malic}},
  \bibinfo {author} {\bibfnamefont {X.}~\bibnamefont {Xu}},\ and\ \bibinfo
  {author} {\bibfnamefont {X.}~\bibnamefont {Li}},\ }\href
  {https://link.aps.org/doi/10.1103/PhysRevB.93.041401} {\bibfield  {journal}
  {\bibinfo  {journal} {Phys. Rev. B}\ }\textbf {\bibinfo {volume} {93}},\
  \bibinfo {pages} {041401} (\bibinfo {year} {2016}{\natexlab{a}})}\BibitemShut
  {NoStop}%
\bibitem [{\citenamefont {Singh}\ \emph
  {et~al.}(2016{\natexlab{b}})\citenamefont {Singh}, \citenamefont {Tran},
  \citenamefont {Kolarczik}, \citenamefont {Seifert}, \citenamefont {Wang},
  \citenamefont {Hao}, \citenamefont {Pleskot}, \citenamefont {Gabor},
  \citenamefont {Helmrich}, \citenamefont {Owschimikow}, \citenamefont
  {Woggon},\ and\ \citenamefont {Li}}]{singhLongLivedValleyPolarization2016}%
  \BibitemOpen
  \bibfield  {author} {\bibinfo {author} {\bibfnamefont {A.}~\bibnamefont
  {Singh}}, \bibinfo {author} {\bibfnamefont {K.}~\bibnamefont {Tran}},
  \bibinfo {author} {\bibfnamefont {M.}~\bibnamefont {Kolarczik}}, \bibinfo
  {author} {\bibfnamefont {J.}~\bibnamefont {Seifert}}, \bibinfo {author}
  {\bibfnamefont {Y.}~\bibnamefont {Wang}}, \bibinfo {author} {\bibfnamefont
  {K.}~\bibnamefont {Hao}}, \bibinfo {author} {\bibfnamefont {D.}~\bibnamefont
  {Pleskot}}, \bibinfo {author} {\bibfnamefont {N.~M.}\ \bibnamefont {Gabor}},
  \bibinfo {author} {\bibfnamefont {S.}~\bibnamefont {Helmrich}}, \bibinfo
  {author} {\bibfnamefont {N.}~\bibnamefont {Owschimikow}}, \bibinfo {author}
  {\bibfnamefont {U.}~\bibnamefont {Woggon}},\ and\ \bibinfo {author}
  {\bibfnamefont {X.}~\bibnamefont {Li}},\ }\href
  {https://link.aps.org/doi/10.1103/PhysRevLett.117.257402} {\bibfield
  {journal} {\bibinfo  {journal} {Phys. Rev. Lett.}\ }\textbf {\bibinfo
  {volume} {117}},\ \bibinfo {pages} {257402} (\bibinfo {year}
  {2016}{\natexlab{b}})}\BibitemShut {NoStop}%
\bibitem [{\citenamefont {An}\ \emph {et~al.}(2023)\citenamefont {An},
  \citenamefont {Soubelet}, \citenamefont {Zhumagulov}, \citenamefont {Zopf},
  \citenamefont {Delhomme}, \citenamefont {Qian}, \citenamefont {Faria~Junior},
  \citenamefont {Fabian}, \citenamefont {Cao}, \citenamefont {Yang},
  \citenamefont {Stier}, \citenamefont {Ding},\ and\ \citenamefont
  {Finley}}]{anStrainControlExciton2023}%
  \BibitemOpen
  \bibfield  {author} {\bibinfo {author} {\bibfnamefont {Z.}~\bibnamefont
  {An}}, \bibinfo {author} {\bibfnamefont {P.}~\bibnamefont {Soubelet}},
  \bibinfo {author} {\bibfnamefont {Y.}~\bibnamefont {Zhumagulov}}, \bibinfo
  {author} {\bibfnamefont {M.}~\bibnamefont {Zopf}}, \bibinfo {author}
  {\bibfnamefont {A.}~\bibnamefont {Delhomme}}, \bibinfo {author}
  {\bibfnamefont {C.}~\bibnamefont {Qian}}, \bibinfo {author} {\bibfnamefont
  {P.~E.}\ \bibnamefont {Faria~Junior}}, \bibinfo {author} {\bibfnamefont
  {J.}~\bibnamefont {Fabian}}, \bibinfo {author} {\bibfnamefont
  {X.}~\bibnamefont {Cao}}, \bibinfo {author} {\bibfnamefont {J.}~\bibnamefont
  {Yang}}, \bibinfo {author} {\bibfnamefont {A.~V.}\ \bibnamefont {Stier}},
  \bibinfo {author} {\bibfnamefont {F.}~\bibnamefont {Ding}},\ and\ \bibinfo
  {author} {\bibfnamefont {J.~J.}\ \bibnamefont {Finley}},\ }\href
  {https://link.aps.org/doi/10.1103/PhysRevB.108.L041404} {\bibfield  {journal}
  {\bibinfo  {journal} {Phys. Rev. B}\ }\textbf {\bibinfo {volume} {108}},\
  \bibinfo {pages} {L041404} (\bibinfo {year} {2023})}\BibitemShut {NoStop}%
\bibitem [{\citenamefont {Plechinger}\ \emph {et~al.}(2016)\citenamefont
  {Plechinger}, \citenamefont {Nagler}, \citenamefont {Arora}, \citenamefont
  {Schmidt}, \citenamefont {Chernikov}, \citenamefont {Del~{\'A}guila},
  \citenamefont {Christianen}, \citenamefont {Bratschitsch}, \citenamefont
  {Sch{\"u}ller},\ and\ \citenamefont
  {Korn}}]{plechingerTrionFineStructure2016}%
  \BibitemOpen
  \bibfield  {author} {\bibinfo {author} {\bibfnamefont {G.}~\bibnamefont
  {Plechinger}}, \bibinfo {author} {\bibfnamefont {P.}~\bibnamefont {Nagler}},
  \bibinfo {author} {\bibfnamefont {A.}~\bibnamefont {Arora}}, \bibinfo
  {author} {\bibfnamefont {R.}~\bibnamefont {Schmidt}}, \bibinfo {author}
  {\bibfnamefont {A.}~\bibnamefont {Chernikov}}, \bibinfo {author}
  {\bibfnamefont {A.~G.}\ \bibnamefont {Del~{\'A}guila}}, \bibinfo {author}
  {\bibfnamefont {P.~C.}\ \bibnamefont {Christianen}}, \bibinfo {author}
  {\bibfnamefont {R.}~\bibnamefont {Bratschitsch}}, \bibinfo {author}
  {\bibfnamefont {C.}~\bibnamefont {Sch{\"u}ller}},\ and\ \bibinfo {author}
  {\bibfnamefont {T.}~\bibnamefont {Korn}},\ }\href
  {https://www.nature.com/articles/ncomms12715} {\bibfield  {journal} {\bibinfo
   {journal} {Nat. Commun.}\ }\textbf {\bibinfo {volume} {7}},\ \bibinfo
  {pages} {12715} (\bibinfo {year} {2016})}\BibitemShut {NoStop}%
\bibitem [{\citenamefont {Beret}\ \emph {et~al.}(2023)\citenamefont {Beret},
  \citenamefont {Ren}, \citenamefont {Robert}, \citenamefont {Foussat},
  \citenamefont {Renucci}, \citenamefont {Lagarde}, \citenamefont {Balocchi},
  \citenamefont {Amand}, \citenamefont {Urbaszek}, \citenamefont {Watanabe},
  \citenamefont {Taniguchi}, \citenamefont {Marie},\ and\ \citenamefont
  {Lombez}}]{beretNonlinearDiffusionNegatively2023}%
  \BibitemOpen
  \bibfield  {author} {\bibinfo {author} {\bibfnamefont {D.}~\bibnamefont
  {Beret}}, \bibinfo {author} {\bibfnamefont {L.}~\bibnamefont {Ren}}, \bibinfo
  {author} {\bibfnamefont {C.}~\bibnamefont {Robert}}, \bibinfo {author}
  {\bibfnamefont {L.}~\bibnamefont {Foussat}}, \bibinfo {author} {\bibfnamefont
  {P.}~\bibnamefont {Renucci}}, \bibinfo {author} {\bibfnamefont
  {D.}~\bibnamefont {Lagarde}}, \bibinfo {author} {\bibfnamefont
  {A.}~\bibnamefont {Balocchi}}, \bibinfo {author} {\bibfnamefont
  {T.}~\bibnamefont {Amand}}, \bibinfo {author} {\bibfnamefont
  {B.}~\bibnamefont {Urbaszek}}, \bibinfo {author} {\bibfnamefont
  {K.}~\bibnamefont {Watanabe}}, \bibinfo {author} {\bibfnamefont
  {T.}~\bibnamefont {Taniguchi}}, \bibinfo {author} {\bibfnamefont
  {X.}~\bibnamefont {Marie}},\ and\ \bibinfo {author} {\bibfnamefont
  {L.}~\bibnamefont {Lombez}},\ }\href
  {https://link.aps.org/doi/10.1103/PhysRevB.107.045420} {\bibfield  {journal}
  {\bibinfo  {journal} {Phys. Rev. B}\ }\textbf {\bibinfo {volume} {107}},\
  \bibinfo {pages} {045420} (\bibinfo {year} {2023})}\BibitemShut {NoStop}%
\bibitem [{\citenamefont {Cheng}\ \emph {et~al.}(2021)\citenamefont {Cheng},
  \citenamefont {Li}, \citenamefont {Jin}, \citenamefont {Zhang},\ and\
  \citenamefont {Wang}}]{chengObservationDiffusionDrift2021}%
  \BibitemOpen
  \bibfield  {author} {\bibinfo {author} {\bibfnamefont {G.}~\bibnamefont
  {Cheng}}, \bibinfo {author} {\bibfnamefont {B.}~\bibnamefont {Li}}, \bibinfo
  {author} {\bibfnamefont {Z.}~\bibnamefont {Jin}}, \bibinfo {author}
  {\bibfnamefont {M.}~\bibnamefont {Zhang}},\ and\ \bibinfo {author}
  {\bibfnamefont {J.}~\bibnamefont {Wang}},\ }\href
  {https://pubs.acs.org/doi/10.1021/acs.nanolett.1c02351} {\bibfield  {journal}
  {\bibinfo  {journal} {Nano Lett.}\ }\textbf {\bibinfo {volume} {21}},\
  \bibinfo {pages} {6314} (\bibinfo {year} {2021})}\BibitemShut {NoStop}%
\bibitem [{\citenamefont {Kim}\ \emph {et~al.}(2022)\citenamefont {Kim},
  \citenamefont {Luo}, \citenamefont {Rhodes}, \citenamefont {Bai},
  \citenamefont {Wang}, \citenamefont {Liu}, \citenamefont {Jordan},
  \citenamefont {Huang}, \citenamefont {Li}, \citenamefont {Taniguchi},
  \citenamefont {Watanabe}, \citenamefont {Owen}, \citenamefont {Strauf},
  \citenamefont {Barmak}, \citenamefont {Zhu},\ and\ \citenamefont
  {Hone}}]{kimFreeTrionsUnity2022}%
  \BibitemOpen
  \bibfield  {author} {\bibinfo {author} {\bibfnamefont {B.}~\bibnamefont
  {Kim}}, \bibinfo {author} {\bibfnamefont {Y.}~\bibnamefont {Luo}}, \bibinfo
  {author} {\bibfnamefont {D.}~\bibnamefont {Rhodes}}, \bibinfo {author}
  {\bibfnamefont {Y.}~\bibnamefont {Bai}}, \bibinfo {author} {\bibfnamefont
  {J.}~\bibnamefont {Wang}}, \bibinfo {author} {\bibfnamefont {S.}~\bibnamefont
  {Liu}}, \bibinfo {author} {\bibfnamefont {A.}~\bibnamefont {Jordan}},
  \bibinfo {author} {\bibfnamefont {B.}~\bibnamefont {Huang}}, \bibinfo
  {author} {\bibfnamefont {Z.}~\bibnamefont {Li}}, \bibinfo {author}
  {\bibfnamefont {T.}~\bibnamefont {Taniguchi}}, \bibinfo {author}
  {\bibfnamefont {K.}~\bibnamefont {Watanabe}}, \bibinfo {author}
  {\bibfnamefont {J.}~\bibnamefont {Owen}}, \bibinfo {author} {\bibfnamefont
  {S.}~\bibnamefont {Strauf}}, \bibinfo {author} {\bibfnamefont
  {K.}~\bibnamefont {Barmak}}, \bibinfo {author} {\bibfnamefont
  {X.}~\bibnamefont {Zhu}},\ and\ \bibinfo {author} {\bibfnamefont
  {J.}~\bibnamefont {Hone}},\ }\href
  {https://pubs.acs.org/doi/10.1021/acsnano.1c04331} {\bibfield  {journal}
  {\bibinfo  {journal} {ACS Nano}\ }\textbf {\bibinfo {volume} {16}},\ \bibinfo
  {pages} {140} (\bibinfo {year} {2022})}\BibitemShut {NoStop}%
\bibitem [{\citenamefont {Qiu}\ \emph {et~al.}(2015)\citenamefont {Qiu},
  \citenamefont {Cao},\ and\ \citenamefont
  {Louie}}]{qiuNonanalyticityValleyQuantum2015}%
  \BibitemOpen
  \bibfield  {author} {\bibinfo {author} {\bibfnamefont {D.~Y.}\ \bibnamefont
  {Qiu}}, \bibinfo {author} {\bibfnamefont {T.}~\bibnamefont {Cao}},\ and\
  \bibinfo {author} {\bibfnamefont {S.~G.}\ \bibnamefont {Louie}},\ }\href
  {https://link.aps.org/doi/10.1103/PhysRevLett.115.176801} {\bibfield
  {journal} {\bibinfo  {journal} {Phys. Rev. Lett.}\ }\textbf {\bibinfo
  {volume} {115}},\ \bibinfo {pages} {176801} (\bibinfo {year}
  {2015})}\BibitemShut {NoStop}%
\bibitem [{\citenamefont {Wu}\ \emph {et~al.}(2015)\citenamefont {Wu},
  \citenamefont {Qu},\ and\ \citenamefont
  {MacDonald}}]{wuExcitonBandStructure2015}%
  \BibitemOpen
  \bibfield  {author} {\bibinfo {author} {\bibfnamefont {F.}~\bibnamefont
  {Wu}}, \bibinfo {author} {\bibfnamefont {F.}~\bibnamefont {Qu}},\ and\
  \bibinfo {author} {\bibfnamefont {A.~H.}\ \bibnamefont {MacDonald}},\ }\href
  {https://link.aps.org/doi/10.1103/PhysRevB.91.075310} {\bibfield  {journal}
  {\bibinfo  {journal} {Phys. Rev. B}\ }\textbf {\bibinfo {volume} {91}},\
  \bibinfo {pages} {075310} (\bibinfo {year} {2015})}\BibitemShut {NoStop}%
\bibitem [{\citenamefont {Shimazaki}\ \emph {et~al.}(2020)\citenamefont
  {Shimazaki}, \citenamefont {Schwartz}, \citenamefont {Watanabe},
  \citenamefont {Taniguchi}, \citenamefont {Kroner},\ and\ \citenamefont
  {Imamo{\u g}lu}}]{shimazakiStronglyCorrelatedElectrons2020}%
  \BibitemOpen
  \bibfield  {author} {\bibinfo {author} {\bibfnamefont {Y.}~\bibnamefont
  {Shimazaki}}, \bibinfo {author} {\bibfnamefont {I.}~\bibnamefont {Schwartz}},
  \bibinfo {author} {\bibfnamefont {K.}~\bibnamefont {Watanabe}}, \bibinfo
  {author} {\bibfnamefont {T.}~\bibnamefont {Taniguchi}}, \bibinfo {author}
  {\bibfnamefont {M.}~\bibnamefont {Kroner}},\ and\ \bibinfo {author}
  {\bibfnamefont {A.}~\bibnamefont {Imamo{\u g}lu}},\ }\href
  {https://www.nature.com/articles/s41586-020-2191-2} {\bibfield  {journal}
  {\bibinfo  {journal} {Nature}\ }\textbf {\bibinfo {volume} {580}},\ \bibinfo
  {pages} {472} (\bibinfo {year} {2020})}\BibitemShut {NoStop}%
\bibitem [{\citenamefont {Simbulan}\ \emph {et~al.}(2021)\citenamefont
  {Simbulan}, \citenamefont {Huang}, \citenamefont {Peng}, \citenamefont {Li},
  \citenamefont {Gomez~Sanchez}, \citenamefont {Lin}, \citenamefont {Lu},
  \citenamefont {Yang}, \citenamefont {Qi}, \citenamefont {Cheng},
  \citenamefont {Lu},\ and\ \citenamefont
  {Lan}}]{simbulanSelectivePhotoexcitationFiniteMomentum2021}%
  \BibitemOpen
  \bibfield  {author} {\bibinfo {author} {\bibfnamefont {K.~B.}\ \bibnamefont
  {Simbulan}}, \bibinfo {author} {\bibfnamefont {T.-D.}\ \bibnamefont {Huang}},
  \bibinfo {author} {\bibfnamefont {G.-H.}\ \bibnamefont {Peng}}, \bibinfo
  {author} {\bibfnamefont {F.}~\bibnamefont {Li}}, \bibinfo {author}
  {\bibfnamefont {O.~J.}\ \bibnamefont {Gomez~Sanchez}}, \bibinfo {author}
  {\bibfnamefont {J.-D.}\ \bibnamefont {Lin}}, \bibinfo {author} {\bibfnamefont
  {C.-I.}\ \bibnamefont {Lu}}, \bibinfo {author} {\bibfnamefont {C.-S.}\
  \bibnamefont {Yang}}, \bibinfo {author} {\bibfnamefont {J.}~\bibnamefont
  {Qi}}, \bibinfo {author} {\bibfnamefont {S.-J.}\ \bibnamefont {Cheng}},
  \bibinfo {author} {\bibfnamefont {T.-H.}\ \bibnamefont {Lu}},\ and\ \bibinfo
  {author} {\bibfnamefont {Y.-W.}\ \bibnamefont {Lan}},\ }\href
  {https://pubs.acs.org/doi/10.1021/acsnano.0c10823} {\bibfield  {journal}
  {\bibinfo  {journal} {ACS Nano}\ }\textbf {\bibinfo {volume} {15}},\ \bibinfo
  {pages} {3481} (\bibinfo {year} {2021})}\BibitemShut {NoStop}%
\bibitem [{\citenamefont {Yu}\ and\ \citenamefont
  {Wu}(2014)}]{yuValleyDepolarizationDue2014}%
  \BibitemOpen
  \bibfield  {author} {\bibinfo {author} {\bibfnamefont {T.}~\bibnamefont
  {Yu}}\ and\ \bibinfo {author} {\bibfnamefont {M.~W.}\ \bibnamefont {Wu}},\
  }\href {https://link.aps.org/doi/10.1103/PhysRevB.89.205303} {\bibfield
  {journal} {\bibinfo  {journal} {Phys. Rev. B}\ }\textbf {\bibinfo {volume}
  {89}},\ \bibinfo {pages} {205303} (\bibinfo {year} {2014})}\BibitemShut
  {NoStop}%
\bibitem [{\citenamefont {Yang}\ \emph {et~al.}(2020)\citenamefont {Yang},
  \citenamefont {Robert}, \citenamefont {Lu}, \citenamefont {Van~Tuan},
  \citenamefont {Smirnov}, \citenamefont {Marie},\ and\ \citenamefont
  {Dery}}]{yangExcitonValleyDepolarization2020}%
  \BibitemOpen
  \bibfield  {author} {\bibinfo {author} {\bibfnamefont {M.}~\bibnamefont
  {Yang}}, \bibinfo {author} {\bibfnamefont {C.}~\bibnamefont {Robert}},
  \bibinfo {author} {\bibfnamefont {Z.}~\bibnamefont {Lu}}, \bibinfo {author}
  {\bibfnamefont {D.}~\bibnamefont {Van~Tuan}}, \bibinfo {author}
  {\bibfnamefont {D.}~\bibnamefont {Smirnov}}, \bibinfo {author} {\bibfnamefont
  {X.}~\bibnamefont {Marie}},\ and\ \bibinfo {author} {\bibfnamefont
  {H.}~\bibnamefont {Dery}},\ }\href
  {https://link.aps.org/doi/10.1103/PhysRevB.101.115307} {\bibfield  {journal}
  {\bibinfo  {journal} {Phys. Rev. B}\ }\textbf {\bibinfo {volume} {101}},\
  \bibinfo {pages} {115307} (\bibinfo {year} {2020})}\BibitemShut {NoStop}%
\bibitem [{\citenamefont {Yu}\ and\ \citenamefont
  {Wu}(2016)}]{yuValleyDepolarizationDynamics2016}%
  \BibitemOpen
  \bibfield  {author} {\bibinfo {author} {\bibfnamefont {T.}~\bibnamefont
  {Yu}}\ and\ \bibinfo {author} {\bibfnamefont {M.~W.}\ \bibnamefont {Wu}},\
  }\href {https://link.aps.org/doi/10.1103/PhysRevB.93.045414} {\bibfield
  {journal} {\bibinfo  {journal} {Phys. Rev. B}\ }\textbf {\bibinfo {volume}
  {93}},\ \bibinfo {pages} {045414} (\bibinfo {year} {2016})}\BibitemShut
  {NoStop}%
\bibitem [{\citenamefont {Wu}\ \emph {et~al.}(2017)\citenamefont {Wu},
  \citenamefont {Lovorn},\ and\ \citenamefont
  {MacDonald}}]{wuTopologicalExcitonBands2017}%
  \BibitemOpen
  \bibfield  {author} {\bibinfo {author} {\bibfnamefont {F.}~\bibnamefont
  {Wu}}, \bibinfo {author} {\bibfnamefont {T.}~\bibnamefont {Lovorn}},\ and\
  \bibinfo {author} {\bibfnamefont {A.~H.}\ \bibnamefont {MacDonald}},\ }\href
  {http://link.aps.org/doi/10.1103/PhysRevLett.118.147401} {\bibfield
  {journal} {\bibinfo  {journal} {Phys. Rev. Lett.}\ }\textbf {\bibinfo
  {volume} {118}},\ \bibinfo {pages} {147401} (\bibinfo {year}
  {2017})}\BibitemShut {NoStop}%
\bibitem [{\citenamefont {Yang}\ \emph {et~al.}(2022)\citenamefont {Yang},
  \citenamefont {Yu},\ and\ \citenamefont
  {Yao}}]{yangChiralExcitonicsMonolayer2022}%
  \BibitemOpen
  \bibfield  {author} {\bibinfo {author} {\bibfnamefont {X.-C.}\ \bibnamefont
  {Yang}}, \bibinfo {author} {\bibfnamefont {H.}~\bibnamefont {Yu}},\ and\
  \bibinfo {author} {\bibfnamefont {W.}~\bibnamefont {Yao}},\ }\href
  {https://link.aps.org/doi/10.1103/PhysRevLett.128.217402} {\bibfield
  {journal} {\bibinfo  {journal} {Phys. Rev. Lett.}\ }\textbf {\bibinfo
  {volume} {128}},\ \bibinfo {pages} {217402} (\bibinfo {year}
  {2022})}\BibitemShut {NoStop}%
\bibitem [{\citenamefont {Zhang}\ \emph {et~al.}(2023)\citenamefont {Zhang},
  \citenamefont {Zhai}, \citenamefont {Deng}, \citenamefont {Yao},\ and\
  \citenamefont {Zhu}}]{zhangSinglePhotonEmitters2023}%
  \BibitemOpen
  \bibfield  {author} {\bibinfo {author} {\bibfnamefont {D.}~\bibnamefont
  {Zhang}}, \bibinfo {author} {\bibfnamefont {D.}~\bibnamefont {Zhai}},
  \bibinfo {author} {\bibfnamefont {S.}~\bibnamefont {Deng}}, \bibinfo {author}
  {\bibfnamefont {W.}~\bibnamefont {Yao}},\ and\ \bibinfo {author}
  {\bibfnamefont {Q.}~\bibnamefont {Zhu}},\ }\href
  {https://pubs.acs.org/doi/10.1021/acs.nanolett.3c00459} {\bibfield  {journal}
  {\bibinfo  {journal} {Nano Lett.}\ }\textbf {\bibinfo {volume} {23}},\
  \bibinfo {pages} {3851} (\bibinfo {year} {2023})}\BibitemShut {NoStop}%
\bibitem [{\citenamefont {Chen}\ \emph {et~al.}(2023)\citenamefont {Chen},
  \citenamefont {Huang},\ and\ \citenamefont
  {Zhu}}]{chenSearchingUnconventionalSuperfluid2023}%
  \BibitemOpen
  \bibfield  {author} {\bibinfo {author} {\bibfnamefont {W.}~\bibnamefont
  {Chen}}, \bibinfo {author} {\bibfnamefont {C.-J.}\ \bibnamefont {Huang}},\
  and\ \bibinfo {author} {\bibfnamefont {Q.}~\bibnamefont {Zhu}},\ }\href
  {https://link.aps.org/doi/10.1103/PhysRevLett.131.236004} {\bibfield
  {journal} {\bibinfo  {journal} {Phys. Rev. Lett.}\ }\textbf {\bibinfo
  {volume} {131}},\ \bibinfo {pages} {236004} (\bibinfo {year}
  {2023})}\BibitemShut {NoStop}%
\bibitem [{\citenamefont {Kane}\ and\ \citenamefont
  {Mele}(2005{\natexlab{a}})}]{kaneQuantumSpinHall2005}%
  \BibitemOpen
  \bibfield  {author} {\bibinfo {author} {\bibfnamefont {C.~L.}\ \bibnamefont
  {Kane}}\ and\ \bibinfo {author} {\bibfnamefont {E.~J.}\ \bibnamefont
  {Mele}},\ }\href {https://link.aps.org/doi/10.1103/PhysRevLett.95.226801}
  {\bibfield  {journal} {\bibinfo  {journal} {Phys. Rev. Lett.}\ }\textbf
  {\bibinfo {volume} {95}},\ \bibinfo {pages} {226801} (\bibinfo {year}
  {2005}{\natexlab{a}})}\BibitemShut {NoStop}%
\bibitem [{\citenamefont {Kane}\ and\ \citenamefont
  {Mele}(2005{\natexlab{b}})}]{kaneTopologicalOrderQuantum2005}%
  \BibitemOpen
  \bibfield  {author} {\bibinfo {author} {\bibfnamefont {C.~L.}\ \bibnamefont
  {Kane}}\ and\ \bibinfo {author} {\bibfnamefont {E.~J.}\ \bibnamefont
  {Mele}},\ }\href {https://link.aps.org/doi/10.1103/PhysRevLett.95.146802}
  {\bibfield  {journal} {\bibinfo  {journal} {Phys. Rev. Lett.}\ }\textbf
  {\bibinfo {volume} {95}},\ \bibinfo {pages} {146802} (\bibinfo {year}
  {2005}{\natexlab{b}})}\BibitemShut {NoStop}%
\bibitem [{\citenamefont {Sauer}\ \emph {et~al.}(2021)\citenamefont {Sauer},
  \citenamefont {Nielsen}, \citenamefont {{Merring-Mikkelsen}},\ and\
  \citenamefont {Pedersen}}]{sauerOpticalEmissionLightlike2021}%
  \BibitemOpen
  \bibfield  {author} {\bibinfo {author} {\bibfnamefont {M.~O.}\ \bibnamefont
  {Sauer}}, \bibinfo {author} {\bibfnamefont {C.~E.~M.}\ \bibnamefont
  {Nielsen}}, \bibinfo {author} {\bibfnamefont {L.}~\bibnamefont
  {{Merring-Mikkelsen}}},\ and\ \bibinfo {author} {\bibfnamefont {T.~G.}\
  \bibnamefont {Pedersen}},\ }\href
  {https://link.aps.org/doi/10.1103/PhysRevB.103.205404} {\bibfield  {journal}
  {\bibinfo  {journal} {Phys. Rev. B}\ }\textbf {\bibinfo {volume} {103}},\
  \bibinfo {pages} {205404} (\bibinfo {year} {2021})}\BibitemShut {NoStop}%
\bibitem [{\citenamefont {Salvador}\ \emph {et~al.}(2022)\citenamefont
  {Salvador}, \citenamefont {Kuhlenkamp}, \citenamefont {Ciorciaro},
  \citenamefont {Knap},\ and\ \citenamefont {{\.I}mamo{\u
  g}lu}}]{salvadorOpticalSignaturesPeriodic2022}%
  \BibitemOpen
  \bibfield  {author} {\bibinfo {author} {\bibfnamefont {A.~G.}\ \bibnamefont
  {Salvador}}, \bibinfo {author} {\bibfnamefont {C.}~\bibnamefont
  {Kuhlenkamp}}, \bibinfo {author} {\bibfnamefont {L.}~\bibnamefont
  {Ciorciaro}}, \bibinfo {author} {\bibfnamefont {M.}~\bibnamefont {Knap}},\
  and\ \bibinfo {author} {\bibfnamefont {A.}~\bibnamefont {{\.I}mamo{\u
  g}lu}},\ }\href {https://link.aps.org/doi/10.1103/PhysRevLett.128.237401}
  {\bibfield  {journal} {\bibinfo  {journal} {Phys. Rev. Lett.}\ }\textbf
  {\bibinfo {volume} {128}},\ \bibinfo {pages} {237401} (\bibinfo {year}
  {2022})}\BibitemShut {NoStop}%
\bibitem [{\citenamefont {Yu}\ \emph {et~al.}(2015{\natexlab{b}})\citenamefont
  {Yu}, \citenamefont {Cui}, \citenamefont {Xu},\ and\ \citenamefont
  {Yao}}]{yuValleyExcitonsTwodimensional2015}%
  \BibitemOpen
  \bibfield  {author} {\bibinfo {author} {\bibfnamefont {H.}~\bibnamefont
  {Yu}}, \bibinfo {author} {\bibfnamefont {X.}~\bibnamefont {Cui}}, \bibinfo
  {author} {\bibfnamefont {X.}~\bibnamefont {Xu}},\ and\ \bibinfo {author}
  {\bibfnamefont {W.}~\bibnamefont {Yao}},\ }\href
  {https://academic.oup.com/nsr/article/2/1/57/2606155} {\bibfield  {journal}
  {\bibinfo  {journal} {Natl. Sci. Rev.}\ }\textbf {\bibinfo {volume} {2}},\
  \bibinfo {pages} {57} (\bibinfo {year} {2015}{\natexlab{b}})}\BibitemShut
  {NoStop}%
\bibitem [{\citenamefont {Hasan}\ and\ \citenamefont
  {Kane}(2010)}]{hasanColloquiumTopologicalInsulators2010}%
  \BibitemOpen
  \bibfield  {author} {\bibinfo {author} {\bibfnamefont {M.~Z.}\ \bibnamefont
  {Hasan}}\ and\ \bibinfo {author} {\bibfnamefont {C.~L.}\ \bibnamefont
  {Kane}},\ }\href {https://link.aps.org/doi/10.1103/RevModPhys.82.3045}
  {\bibfield  {journal} {\bibinfo  {journal} {Rev. Mod. Phys.}\ }\textbf
  {\bibinfo {volume} {82}},\ \bibinfo {pages} {3045} (\bibinfo {year}
  {2010})}\BibitemShut {NoStop}%
\bibitem [{\citenamefont {Zhao}\ \emph {et~al.}(2021)\citenamefont {Zhao},
  \citenamefont {Xiao},\ and\ \citenamefont
  {Yao}}]{zhaoUniversalSuperlatticePotential2021}%
  \BibitemOpen
  \bibfield  {author} {\bibinfo {author} {\bibfnamefont {P.}~\bibnamefont
  {Zhao}}, \bibinfo {author} {\bibfnamefont {C.}~\bibnamefont {Xiao}},\ and\
  \bibinfo {author} {\bibfnamefont {W.}~\bibnamefont {Yao}},\ }\href
  {https://www.nature.com/articles/s41699-021-00221-4} {\bibfield  {journal}
  {\bibinfo  {journal} {npj 2D Mater. Appl.}\ }\textbf {\bibinfo {volume}
  {5}},\ \bibinfo {pages} {38} (\bibinfo {year} {2021})}\BibitemShut {NoStop}%
\bibitem [{\citenamefont {Kim}\ \emph {et~al.}(2024)\citenamefont {Kim},
  \citenamefont {Dominguez}, \citenamefont {{Mayorga-Luna}}, \citenamefont
  {Ye}, \citenamefont {Embley}, \citenamefont {Tan}, \citenamefont {Ni},
  \citenamefont {Liu}, \citenamefont {Ford}, \citenamefont {Gao}, \citenamefont
  {Arash}, \citenamefont {Watanabe}, \citenamefont {Taniguchi}, \citenamefont
  {Kim}, \citenamefont {Shih}, \citenamefont {Lai}, \citenamefont {Yao},
  \citenamefont {Yang}, \citenamefont {Li},\ and\ \citenamefont
  {Miyahara}}]{kimElectrostaticMoirePotential2024}%
  \BibitemOpen
  \bibfield  {author} {\bibinfo {author} {\bibfnamefont {D.~S.}\ \bibnamefont
  {Kim}}, \bibinfo {author} {\bibfnamefont {R.~C.}\ \bibnamefont {Dominguez}},
  \bibinfo {author} {\bibfnamefont {R.}~\bibnamefont {{Mayorga-Luna}}},
  \bibinfo {author} {\bibfnamefont {D.}~\bibnamefont {Ye}}, \bibinfo {author}
  {\bibfnamefont {J.}~\bibnamefont {Embley}}, \bibinfo {author} {\bibfnamefont
  {T.}~\bibnamefont {Tan}}, \bibinfo {author} {\bibfnamefont {Y.}~\bibnamefont
  {Ni}}, \bibinfo {author} {\bibfnamefont {Z.}~\bibnamefont {Liu}}, \bibinfo
  {author} {\bibfnamefont {M.}~\bibnamefont {Ford}}, \bibinfo {author}
  {\bibfnamefont {F.~Y.}\ \bibnamefont {Gao}}, \bibinfo {author} {\bibfnamefont
  {S.}~\bibnamefont {Arash}}, \bibinfo {author} {\bibfnamefont
  {K.}~\bibnamefont {Watanabe}}, \bibinfo {author} {\bibfnamefont
  {T.}~\bibnamefont {Taniguchi}}, \bibinfo {author} {\bibfnamefont
  {S.}~\bibnamefont {Kim}}, \bibinfo {author} {\bibfnamefont {C.-K.}\
  \bibnamefont {Shih}}, \bibinfo {author} {\bibfnamefont {K.}~\bibnamefont
  {Lai}}, \bibinfo {author} {\bibfnamefont {W.}~\bibnamefont {Yao}}, \bibinfo
  {author} {\bibfnamefont {L.}~\bibnamefont {Yang}}, \bibinfo {author}
  {\bibfnamefont {X.}~\bibnamefont {Li}},\ and\ \bibinfo {author}
  {\bibfnamefont {Y.}~\bibnamefont {Miyahara}},\ }\href
  {https://doi.org/10.1038/s41563-023-01637-7} {\bibfield  {journal} {\bibinfo
  {journal} {Nat. Mater.}\ }\textbf {\bibinfo {volume} {23}},\ \bibinfo {pages}
  {65} (\bibinfo {year} {2024})}\BibitemShut {NoStop}%
\bibitem [{\citenamefont {Wang}\ \emph {et~al.}(2025)\citenamefont {Wang},
  \citenamefont {Xu}, \citenamefont {Aronson}, \citenamefont {Bennett},
  \citenamefont {Paul}, \citenamefont {Crowley}, \citenamefont {Collignon},
  \citenamefont {Watanabe}, \citenamefont {Taniguchi}, \citenamefont {Ashoori},
  \citenamefont {Kaxiras}, \citenamefont {Zhang}, \citenamefont
  {{Jarillo-Herrero}},\ and\ \citenamefont
  {Yasuda}}]{wangMoireBandStructure2025}%
  \BibitemOpen
  \bibfield  {author} {\bibinfo {author} {\bibfnamefont {X.}~\bibnamefont
  {Wang}}, \bibinfo {author} {\bibfnamefont {C.}~\bibnamefont {Xu}}, \bibinfo
  {author} {\bibfnamefont {S.}~\bibnamefont {Aronson}}, \bibinfo {author}
  {\bibfnamefont {D.}~\bibnamefont {Bennett}}, \bibinfo {author} {\bibfnamefont
  {N.}~\bibnamefont {Paul}}, \bibinfo {author} {\bibfnamefont {P.~J.~D.}\
  \bibnamefont {Crowley}}, \bibinfo {author} {\bibfnamefont {C.}~\bibnamefont
  {Collignon}}, \bibinfo {author} {\bibfnamefont {K.}~\bibnamefont {Watanabe}},
  \bibinfo {author} {\bibfnamefont {T.}~\bibnamefont {Taniguchi}}, \bibinfo
  {author} {\bibfnamefont {R.}~\bibnamefont {Ashoori}}, \bibinfo {author}
  {\bibfnamefont {E.}~\bibnamefont {Kaxiras}}, \bibinfo {author} {\bibfnamefont
  {Y.}~\bibnamefont {Zhang}}, \bibinfo {author} {\bibfnamefont
  {P.}~\bibnamefont {{Jarillo-Herrero}}},\ and\ \bibinfo {author}
  {\bibfnamefont {K.}~\bibnamefont {Yasuda}},\ }\href
  {https://www.nature.com/articles/s41467-024-55432-2} {\bibfield  {journal}
  {\bibinfo  {journal} {Nat. Commun.}\ }\textbf {\bibinfo {volume} {16}},\
  \bibinfo {pages} {178} (\bibinfo {year} {2025})}\BibitemShut {NoStop}%
\bibitem [{\citenamefont {Kiper}\ \emph {et~al.}(2025)\citenamefont {Kiper},
  \citenamefont {Adlong}, \citenamefont {Christianen}, \citenamefont {Kroner},
  \citenamefont {Watanabe}, \citenamefont {Taniguchi},\ and\ \citenamefont
  {{\.I}mamo{\u g}lu}}]{kiperConfinedTrionsMottWigner2025}%
  \BibitemOpen
  \bibfield  {author} {\bibinfo {author} {\bibfnamefont {N.}~\bibnamefont
  {Kiper}}, \bibinfo {author} {\bibfnamefont {H.~S.}\ \bibnamefont {Adlong}},
  \bibinfo {author} {\bibfnamefont {A.}~\bibnamefont {Christianen}}, \bibinfo
  {author} {\bibfnamefont {M.}~\bibnamefont {Kroner}}, \bibinfo {author}
  {\bibfnamefont {K.}~\bibnamefont {Watanabe}}, \bibinfo {author}
  {\bibfnamefont {T.}~\bibnamefont {Taniguchi}},\ and\ \bibinfo {author}
  {\bibfnamefont {A.}~\bibnamefont {{\.I}mamo{\u g}lu}},\ }\href
  {https://link.aps.org/doi/10.1103/PhysRevX.15.011049} {\bibfield  {journal}
  {\bibinfo  {journal} {Phys. Rev. X}\ }\textbf {\bibinfo {volume} {15}},\
  \bibinfo {pages} {011049} (\bibinfo {year} {2025})}\BibitemShut {NoStop}%
\bibitem [{\citenamefont {Stier}\ \emph {et~al.}(2018)\citenamefont {Stier},
  \citenamefont {Wilson}, \citenamefont {Velizhanin}, \citenamefont {Kono},
  \citenamefont {Xu},\ and\ \citenamefont
  {Crooker}}]{stierMagnetoopticsExcitonRydberg2018}%
  \BibitemOpen
  \bibfield  {author} {\bibinfo {author} {\bibfnamefont {A.~V.}\ \bibnamefont
  {Stier}}, \bibinfo {author} {\bibfnamefont {N.~P.}\ \bibnamefont {Wilson}},
  \bibinfo {author} {\bibfnamefont {K.~A.}\ \bibnamefont {Velizhanin}},
  \bibinfo {author} {\bibfnamefont {J.}~\bibnamefont {Kono}}, \bibinfo {author}
  {\bibfnamefont {X.}~\bibnamefont {Xu}},\ and\ \bibinfo {author}
  {\bibfnamefont {S.~A.}\ \bibnamefont {Crooker}},\ }\href
  {https://link.aps.org/doi/10.1103/PhysRevLett.120.057405} {\bibfield
  {journal} {\bibinfo  {journal} {Phys. Rev. Lett.}\ }\textbf {\bibinfo
  {volume} {120}},\ \bibinfo {pages} {057405} (\bibinfo {year}
  {2018})}\BibitemShut {NoStop}%
\bibitem [{\citenamefont {Wang}\ \emph {et~al.}(2018)\citenamefont {Wang},
  \citenamefont {Chernikov}, \citenamefont {Glazov}, \citenamefont {Heinz},
  \citenamefont {Marie}, \citenamefont {Amand},\ and\ \citenamefont
  {Urbaszek}}]{wangColloquiumExcitonsAtomically2018}%
  \BibitemOpen
  \bibfield  {author} {\bibinfo {author} {\bibfnamefont {G.}~\bibnamefont
  {Wang}}, \bibinfo {author} {\bibfnamefont {A.}~\bibnamefont {Chernikov}},
  \bibinfo {author} {\bibfnamefont {M.~M.}\ \bibnamefont {Glazov}}, \bibinfo
  {author} {\bibfnamefont {T.~F.}\ \bibnamefont {Heinz}}, \bibinfo {author}
  {\bibfnamefont {X.}~\bibnamefont {Marie}}, \bibinfo {author} {\bibfnamefont
  {T.}~\bibnamefont {Amand}},\ and\ \bibinfo {author} {\bibfnamefont
  {B.}~\bibnamefont {Urbaszek}},\ }\href
  {https://link.aps.org/doi/10.1103/RevModPhys.90.021001} {\bibfield  {journal}
  {\bibinfo  {journal} {Rev. Mod. Phys.}\ }\textbf {\bibinfo {volume} {90}},\
  \bibinfo {pages} {021001} (\bibinfo {year} {2018})}\BibitemShut {NoStop}%
\bibitem [{\citenamefont {Liu}\ \emph {et~al.}(2021)\citenamefont {Liu},
  \citenamefont {Barr{\'e}}, \citenamefont {Van~Baren}, \citenamefont {Wilson},
  \citenamefont {Taniguchi}, \citenamefont {Watanabe}, \citenamefont {Cui},
  \citenamefont {Gabor}, \citenamefont {Heinz}, \citenamefont {Chang},\ and\
  \citenamefont {Lui}}]{liuSignaturesMoireTrions2021}%
  \BibitemOpen
  \bibfield  {author} {\bibinfo {author} {\bibfnamefont {E.}~\bibnamefont
  {Liu}}, \bibinfo {author} {\bibfnamefont {E.}~\bibnamefont {Barr{\'e}}},
  \bibinfo {author} {\bibfnamefont {J.}~\bibnamefont {Van~Baren}}, \bibinfo
  {author} {\bibfnamefont {M.}~\bibnamefont {Wilson}}, \bibinfo {author}
  {\bibfnamefont {T.}~\bibnamefont {Taniguchi}}, \bibinfo {author}
  {\bibfnamefont {K.}~\bibnamefont {Watanabe}}, \bibinfo {author}
  {\bibfnamefont {Y.-T.}\ \bibnamefont {Cui}}, \bibinfo {author} {\bibfnamefont
  {N.~M.}\ \bibnamefont {Gabor}}, \bibinfo {author} {\bibfnamefont {T.~F.}\
  \bibnamefont {Heinz}}, \bibinfo {author} {\bibfnamefont {Y.-C.}\ \bibnamefont
  {Chang}},\ and\ \bibinfo {author} {\bibfnamefont {C.~H.}\ \bibnamefont
  {Lui}},\ }\href {https://www.nature.com/articles/s41586-021-03541-z}
  {\bibfield  {journal} {\bibinfo  {journal} {Nature}\ }\textbf {\bibinfo
  {volume} {594}},\ \bibinfo {pages} {46} (\bibinfo {year} {2021})}\BibitemShut
  {NoStop}%
\bibitem [{\citenamefont {Wang}\ \emph {et~al.}(2021)\citenamefont {Wang},
  \citenamefont {Zhu}, \citenamefont {Seyler}, \citenamefont {Rivera},
  \citenamefont {Zheng}, \citenamefont {Wang}, \citenamefont {He},
  \citenamefont {Taniguchi}, \citenamefont {Watanabe}, \citenamefont {Yan},
  \citenamefont {Mandrus}, \citenamefont {Gamelin}, \citenamefont {Yao},\ and\
  \citenamefont {Xu}}]{wangMoireTrionsMoSe22021}%
  \BibitemOpen
  \bibfield  {author} {\bibinfo {author} {\bibfnamefont {X.}~\bibnamefont
  {Wang}}, \bibinfo {author} {\bibfnamefont {J.}~\bibnamefont {Zhu}}, \bibinfo
  {author} {\bibfnamefont {K.~L.}\ \bibnamefont {Seyler}}, \bibinfo {author}
  {\bibfnamefont {P.}~\bibnamefont {Rivera}}, \bibinfo {author} {\bibfnamefont
  {H.}~\bibnamefont {Zheng}}, \bibinfo {author} {\bibfnamefont
  {Y.}~\bibnamefont {Wang}}, \bibinfo {author} {\bibfnamefont {M.}~\bibnamefont
  {He}}, \bibinfo {author} {\bibfnamefont {T.}~\bibnamefont {Taniguchi}},
  \bibinfo {author} {\bibfnamefont {K.}~\bibnamefont {Watanabe}}, \bibinfo
  {author} {\bibfnamefont {J.}~\bibnamefont {Yan}}, \bibinfo {author}
  {\bibfnamefont {D.~G.}\ \bibnamefont {Mandrus}}, \bibinfo {author}
  {\bibfnamefont {D.~R.}\ \bibnamefont {Gamelin}}, \bibinfo {author}
  {\bibfnamefont {W.}~\bibnamefont {Yao}},\ and\ \bibinfo {author}
  {\bibfnamefont {X.}~\bibnamefont {Xu}},\ }\href
  {https://www.nature.com/articles/s41565-021-00969-2} {\bibfield  {journal}
  {\bibinfo  {journal} {Nat. Nanotechnol.}\ }\textbf {\bibinfo {volume} {16}},\
  \bibinfo {pages} {1208} (\bibinfo {year} {2021})}\BibitemShut {NoStop}%
\bibitem [{\citenamefont {{Brotons-Gisbert}}\ \emph {et~al.}(2021)\citenamefont
  {{Brotons-Gisbert}}, \citenamefont {Baek}, \citenamefont {Campbell},
  \citenamefont {Watanabe}, \citenamefont {Taniguchi},\ and\ \citenamefont
  {Gerardot}}]{brotons-gisbertMoireTrappedInterlayerTrions2021}%
  \BibitemOpen
  \bibfield  {author} {\bibinfo {author} {\bibfnamefont {M.}~\bibnamefont
  {{Brotons-Gisbert}}}, \bibinfo {author} {\bibfnamefont {H.}~\bibnamefont
  {Baek}}, \bibinfo {author} {\bibfnamefont {A.}~\bibnamefont {Campbell}},
  \bibinfo {author} {\bibfnamefont {K.}~\bibnamefont {Watanabe}}, \bibinfo
  {author} {\bibfnamefont {T.}~\bibnamefont {Taniguchi}},\ and\ \bibinfo
  {author} {\bibfnamefont {B.~D.}\ \bibnamefont {Gerardot}},\ }\href
  {https://link.aps.org/doi/10.1103/PhysRevX.11.031033} {\bibfield  {journal}
  {\bibinfo  {journal} {Phys. Rev. X}\ }\textbf {\bibinfo {volume} {11}},\
  \bibinfo {pages} {031033} (\bibinfo {year} {2021})}\BibitemShut {NoStop}%
\bibitem [{\citenamefont {Marcellina}\ \emph {et~al.}(2021)\citenamefont
  {Marcellina}, \citenamefont {Liu}, \citenamefont {Hu}, \citenamefont
  {Fieramosca}, \citenamefont {Huang}, \citenamefont {Du}, \citenamefont {Liu},
  \citenamefont {Zhao}, \citenamefont {Watanabe}, \citenamefont {Taniguchi},\
  and\ \citenamefont {Xiong}}]{marcellinaEvidenceMoireTrions2021}%
  \BibitemOpen
  \bibfield  {author} {\bibinfo {author} {\bibfnamefont {E.}~\bibnamefont
  {Marcellina}}, \bibinfo {author} {\bibfnamefont {X.}~\bibnamefont {Liu}},
  \bibinfo {author} {\bibfnamefont {Z.}~\bibnamefont {Hu}}, \bibinfo {author}
  {\bibfnamefont {A.}~\bibnamefont {Fieramosca}}, \bibinfo {author}
  {\bibfnamefont {Y.}~\bibnamefont {Huang}}, \bibinfo {author} {\bibfnamefont
  {W.}~\bibnamefont {Du}}, \bibinfo {author} {\bibfnamefont {S.}~\bibnamefont
  {Liu}}, \bibinfo {author} {\bibfnamefont {J.}~\bibnamefont {Zhao}}, \bibinfo
  {author} {\bibfnamefont {K.}~\bibnamefont {Watanabe}}, \bibinfo {author}
  {\bibfnamefont {T.}~\bibnamefont {Taniguchi}},\ and\ \bibinfo {author}
  {\bibfnamefont {Q.}~\bibnamefont {Xiong}},\ }\href
  {https://pubs.acs.org/doi/10.1021/acs.nanolett.1c01207} {\bibfield  {journal}
  {\bibinfo  {journal} {Nano Lett.}\ }\textbf {\bibinfo {volume} {21}},\
  \bibinfo {pages} {4461} (\bibinfo {year} {2021})}\BibitemShut {NoStop}%
\bibitem [{\citenamefont {K{\"o}nig}\ \emph {et~al.}(2007)\citenamefont
  {K{\"o}nig}, \citenamefont {Wiedmann}, \citenamefont {Br{\"u}ne},
  \citenamefont {Roth}, \citenamefont {Buhmann}, \citenamefont {Molenkamp},
  \citenamefont {Qi},\ and\ \citenamefont {Zhang}}]{konigQuantumSpinHall2007}%
  \BibitemOpen
  \bibfield  {author} {\bibinfo {author} {\bibfnamefont {M.}~\bibnamefont
  {K{\"o}nig}}, \bibinfo {author} {\bibfnamefont {S.}~\bibnamefont {Wiedmann}},
  \bibinfo {author} {\bibfnamefont {C.}~\bibnamefont {Br{\"u}ne}}, \bibinfo
  {author} {\bibfnamefont {A.}~\bibnamefont {Roth}}, \bibinfo {author}
  {\bibfnamefont {H.}~\bibnamefont {Buhmann}}, \bibinfo {author} {\bibfnamefont
  {L.~W.}\ \bibnamefont {Molenkamp}}, \bibinfo {author} {\bibfnamefont {X.-L.}\
  \bibnamefont {Qi}},\ and\ \bibinfo {author} {\bibfnamefont {S.-C.}\
  \bibnamefont {Zhang}},\ }\href
  {https://www.science.org/doi/10.1126/science.1148047} {\bibfield  {journal}
  {\bibinfo  {journal} {Science}\ }\textbf {\bibinfo {volume} {318}},\ \bibinfo
  {pages} {766} (\bibinfo {year} {2007})}\BibitemShut {NoStop}%
\bibitem [{\citenamefont {Bernevig}\ \emph {et~al.}(2006)\citenamefont
  {Bernevig}, \citenamefont {Hughes},\ and\ \citenamefont
  {Zhang}}]{bernevigQuantumSpinHall2006}%
  \BibitemOpen
  \bibfield  {author} {\bibinfo {author} {\bibfnamefont {B.~A.}\ \bibnamefont
  {Bernevig}}, \bibinfo {author} {\bibfnamefont {T.~L.}\ \bibnamefont
  {Hughes}},\ and\ \bibinfo {author} {\bibfnamefont {S.-C.}\ \bibnamefont
  {Zhang}},\ }\href {https://www.science.org/doi/10.1126/science.1133734}
  {\bibfield  {journal} {\bibinfo  {journal} {Science}\ }\textbf {\bibinfo
  {volume} {314}},\ \bibinfo {pages} {1757} (\bibinfo {year}
  {2006})}\BibitemShut {NoStop}%
\bibitem [{\citenamefont {Onga}\ \emph {et~al.}(2017)\citenamefont {Onga},
  \citenamefont {Zhang}, \citenamefont {Ideue},\ and\ \citenamefont
  {Iwasa}}]{ongaExcitonHallEffect2017}%
  \BibitemOpen
  \bibfield  {author} {\bibinfo {author} {\bibfnamefont {M.}~\bibnamefont
  {Onga}}, \bibinfo {author} {\bibfnamefont {Y.}~\bibnamefont {Zhang}},
  \bibinfo {author} {\bibfnamefont {T.}~\bibnamefont {Ideue}},\ and\ \bibinfo
  {author} {\bibfnamefont {Y.}~\bibnamefont {Iwasa}},\ }\href
  {https://www.nature.com/articles/nmat4996} {\bibfield  {journal} {\bibinfo
  {journal} {Nat. Mater.}\ }\textbf {\bibinfo {volume} {16}},\ \bibinfo {pages}
  {1193} (\bibinfo {year} {2017})}\BibitemShut {NoStop}%
\bibitem [{\citenamefont {Huang}\ \emph {et~al.}(2020)\citenamefont {Huang},
  \citenamefont {Liu}, \citenamefont {Dini}, \citenamefont {Tan}, \citenamefont
  {Liu}, \citenamefont {Fang}, \citenamefont {Liu}, \citenamefont {Liew},\ and\
  \citenamefont {Gao}}]{huangRobustRoomTemperature2020}%
  \BibitemOpen
  \bibfield  {author} {\bibinfo {author} {\bibfnamefont {Z.}~\bibnamefont
  {Huang}}, \bibinfo {author} {\bibfnamefont {Y.}~\bibnamefont {Liu}}, \bibinfo
  {author} {\bibfnamefont {K.}~\bibnamefont {Dini}}, \bibinfo {author}
  {\bibfnamefont {Q.}~\bibnamefont {Tan}}, \bibinfo {author} {\bibfnamefont
  {Z.}~\bibnamefont {Liu}}, \bibinfo {author} {\bibfnamefont {H.}~\bibnamefont
  {Fang}}, \bibinfo {author} {\bibfnamefont {J.}~\bibnamefont {Liu}}, \bibinfo
  {author} {\bibfnamefont {T.}~\bibnamefont {Liew}},\ and\ \bibinfo {author}
  {\bibfnamefont {W.}~\bibnamefont {Gao}},\ }\href
  {https://pubs.acs.org/doi/10.1021/acs.nanolett.9b04836} {\bibfield  {journal}
  {\bibinfo  {journal} {Nano Lett.}\ }\textbf {\bibinfo {volume} {20}},\
  \bibinfo {pages} {1345} (\bibinfo {year} {2020})}\BibitemShut {NoStop}%
\bibitem [{\citenamefont {Liang}\ \emph {et~al.}(2017)\citenamefont {Liang},
  \citenamefont {Yang}, \citenamefont {Renucci}, \citenamefont {Tao},
  \citenamefont {Laczkowski}, \citenamefont {{Mc-Murtry}}, \citenamefont
  {Wang}, \citenamefont {Marie}, \citenamefont {George}, \citenamefont
  {{Petit-Watelot}}, \citenamefont {Djeffal}, \citenamefont {Mangin},
  \citenamefont {Jaffr{\`e}s},\ and\ \citenamefont
  {Lu}}]{liangElectricalSpinInjection2017}%
  \BibitemOpen
  \bibfield  {author} {\bibinfo {author} {\bibfnamefont {S.}~\bibnamefont
  {Liang}}, \bibinfo {author} {\bibfnamefont {H.}~\bibnamefont {Yang}},
  \bibinfo {author} {\bibfnamefont {P.}~\bibnamefont {Renucci}}, \bibinfo
  {author} {\bibfnamefont {B.}~\bibnamefont {Tao}}, \bibinfo {author}
  {\bibfnamefont {P.}~\bibnamefont {Laczkowski}}, \bibinfo {author}
  {\bibfnamefont {S.}~\bibnamefont {{Mc-Murtry}}}, \bibinfo {author}
  {\bibfnamefont {G.}~\bibnamefont {Wang}}, \bibinfo {author} {\bibfnamefont
  {X.}~\bibnamefont {Marie}}, \bibinfo {author} {\bibfnamefont {J.-M.}\
  \bibnamefont {George}}, \bibinfo {author} {\bibfnamefont {S.}~\bibnamefont
  {{Petit-Watelot}}}, \bibinfo {author} {\bibfnamefont {A.}~\bibnamefont
  {Djeffal}}, \bibinfo {author} {\bibfnamefont {S.}~\bibnamefont {Mangin}},
  \bibinfo {author} {\bibfnamefont {H.}~\bibnamefont {Jaffr{\`e}s}},\ and\
  \bibinfo {author} {\bibfnamefont {Y.}~\bibnamefont {Lu}},\ }\href
  {https://www.nature.com/articles/ncomms14947} {\bibfield  {journal} {\bibinfo
   {journal} {Nat. Commun.}\ }\textbf {\bibinfo {volume} {8}},\ \bibinfo
  {pages} {14947} (\bibinfo {year} {2017})}\BibitemShut {NoStop}%
\end{thebibliography}
\end{document}